\documentclass[journal=jpc,manuscript=article]{achemso}
\usepackage{longtable}
\usepackage{multirow}
\usepackage{amsmath}
\usepackage{subfigure}
\usepackage[usenames,dvipsnames]{xcolor}
\usepackage{soul}
\usepackage{bm}

\usepackage{xr}
\usepackage{amssymb}
\usepackage{siunitx}
\usepackage{multicol} \usepackage{wrapfig} \usepackage{enumitem}
\usepackage{bm} \usepackage{outlines} 
\usepackage[normalem]{ulem}
\usepackage{url}
\usepackage{booktabs}
\usepackage{graphicx}
\usepackage[normalem]{ulem}
\usepackage{hyperref}
\usepackage{physics}
\usepackage{siunitx}\usepackage{booktabs}

\makeatletter
\newcommand*{\addFileDependency}[1]{
  \typeout{(#1)}
  \@addtofilelist{#1}
  \IfFileExists{#1}{}{\typeout{No file #1.}}
}
\makeatother

\newcommand*{\myexternaldocument}[1]{%
    \externaldocument{#1}%
    \addFileDependency{#1.tex}%
    \addFileDependency{#1.aux}%
}

\myexternaldocument{si}

\makeatletter

\SectionNumbersOn

\author{Meenu Upadhyay} \affiliation[University of Basel]{Department
  of Chemistry, University of Basel, Klingelbergstrasse 80, CH-4056
  Basel, Switzerland.}

\author{Markus Meuwly} \affiliation[University of Basel]{Department of
  Chemistry, University of Basel, Klingelbergstrasse 80 , CH-4056
  Basel, Switzerland.}  \email{m.meuwly@unibas.ch}

\title{CO$_2$ and NO$_2$ Formation on Amorphous Solid Water}

\begin{document}

\date{\today}

\begin{abstract}
\noindent
 The dynamics for molecule formation, relaxation, diffusion, and desorption on amorphous solid water is studied in a quantitative fashion. We aim at characterizing, at a quantitative level, the formation probability, stabilization, energy relaxation and diffusion dynamics of CO$_2$ and NO$_2$ on cold amorphous solid water following atom+diatom recombination reactions. Accurate machine-learned energy functions combined with fluctuating charge models were used to investigate the diffusion, interactions, and recombination dynamics of atomic oxygen with CO and NO on amorphous solid water (ASW). Energy relaxation to the ASW and into water-internal-degrees of freedom were determined from analysis of the vibrational density of states. The surface diffusion and desorption energetics was investigated from extended and nonequilibrium MD simulations. The reaction probability on the nanosecond time scale is determined in a quantitative fashion and demonstrates that surface diffusion of the reactants leads to recombination for initial separations up to 20 \AA\/. After recombination both, CO$_2$ and NO$_2$, stabilize by energy transfer to water internal and surface phonon modes on the picosecond time scale. The average diffusion barriers and desorption energies agree with those reported from experiments. After recombination, the triatomic products diffuse easily which contrasts with the equilibrium situation in which both, CO$_2$ and NO$_2$, are stationary on the multi-nanosecond time scale. 
\end{abstract}

\section{Introduction} 
Surface processes are of paramount importance for the genesis of
molecules in the universe. In the interstellar medium, nitrogen
emerges as one of the chemically dynamic species, following hydrogen,
oxygen, and carbon. Within the realm of prebiotic molecules and simple
amino acids including CH$_2$NH, CH$_3$NH$_2$, NH$_2$CH$_2$COOH,
NH$_2$CHO, HNCO, nitrogen stands out as the common element,
underscoring the necessity of exploring the reactivity of
nitrogen-containing molecules. This is because three-body collisions
in the gas phase are highly inefficient for molecule formation from
atomic constituents. Astrophysically relevant surfaces consist of
silicates, carbonaceous species (graphite, amorphous carbon,
polyaromatic hydrocarbons), or water.\cite{herbst:2001} In cold
molecular clouds, cosmic dust grains are typically covered by water
ice which can be polycrystalline
or amorphous (ASW). At low
temperatures, ASW dominates over the polycrystalline
phase.\cite{wakelam:2017} Typically, bulk water is present in the
form of ASW which is the main component of interstellar
ices.\cite{hagen:1981} The structure of ASW is usually probed by
spectroscopic measurements\cite{hagen:1981,jenniskens:1994} although
interference-based methods have also been
employed.\cite{linnartz:2012} ASWs are porous structures
characterized by surface roughness and internal cavities of different
sizes which can retain molecular or atomic guests.\cite{barnun:1987}
Under laboratory conditions the water ices have been reported to be
porous\cite{he:2016,kouchi:2020} or
non-porous\cite{Oba09p464,he:2016,kouchi:2020} ASW whereas the
morphology of ices in the interstellar medium is more heavily
debated.\cite{keane:2001,kouchi:2021}\\

\noindent
The high porosity of ASW\cite{bossa:2014,bossa:2015,cazauxs:2015}
makes it a suitable catalyst for gas-surface reactions involving
oxygen\cite{Ioppolo:2011,romanzin:2011,Chaabouni:2012,oxy.diff.minissale:2013,o2.dulieu:2016,MM.oxy:2018,MM.oxy:2019,christianson:2021},
hydrogen\cite{hama:2013},
carbonaceous\cite{minissale:2013,minissale:2016,qasim:2020,molpeceres:2021}
and nitrogen-containing\cite{no1.minissale:2014} species. The surface
morphology and chemistry help to adsorb chemical reagents on top
of\cite{minissale:2018} or inside ASW.\cite{dulieu:2016,tsuge:2020}
This increases the probability for the reaction partners to diffuse to
locations for collisions and association reactions to occur. As the
diffusivity of individual atoms and small molecules has been
established from both, experiments and
simulations,\cite{minissale:2013,MM.oxy:2014,MM.oxy:2018} this is a
likely scenario for formation of molecules on and within ASW.\\

\noindent
The chemical precursors for formation of CO$_2$ are believed to be
carbon monoxide and atomic oxygen and the CO+O reaction has been
proposed as a non-energetic pathway, close to conditions in
interstellar environments, for CO$_2$ formation 20 years ago from
experiments involving a water-ice cap on top of CO and O deposited on
a copper surface.\cite{roser:2001} Formation of CO$_2$($^1
\Sigma_{\rm g}^{+}$) from ground state CO($^1 \Sigma^+$) and
electronically excited O($^1$D) is barrierless. In addition to CO$_2$,
nitrogen-containing species, such as nitric oxide
(NO)\cite{liszt_NO1978,ligterink2018alma,codella2018nitrogen,ziurys1991nitric,mcgonagle1990detection},
nitrous oxide (N$_2$O)\cite{ziurys1994detection,ligterink2018alma}
and nitrosyl hydride (HNO)\cite{snyder1993new} have been detected in
the interstellar medium. However, their interstellar chemistry has
been little explored so far. Specifically, NO is believed to be
critical for the overall nitrogen chemistry of the interstellar
medium.\cite{de2016formation,congiu2012no}\\

\noindent
The quest of the present work is to investigate the energetics and
dynamics of the CO and NO oxygenation reactions O($^1$D)+CO($^1
\Sigma^+$) and O($^3$P) + NO(X$^2\Pi $) on ASW to form ground state
CO$_2$($^1 \Sigma_{\rm g}^{+}$) and NO$_2$($^2$A'), respectively, at
conditions representative of interstellar environments. Electronically
excited atomic oxygen species can, for example, be generated from
photolysis of H$_2$O\cite{klemm:1975} which has a radiative lifetime
of 110 minutes.\cite{garstang:1951} An alternative pathway proceeds
via electron-induced neutral dissociation of water into H$_2$ +
O($^1$D).\cite{schmidt:2019} In the presence of CO, formation of
CO$_2$ in cryogenic CO/H$_2$O films was observed.\cite{schmidt:2019}
Ground state $^3$P atomic oxygen was observed from photolysis of O$_2$
using far-ultraviolet light.\cite{ogilvie:2018} These are also
formation routes of O($^3$P / $^1$D) that can occur in interstellar
photon dominated regions.\\

\noindent
The present work first describes the methods used. This is followed by
CO$_2$ and NO$_2$ formation dynamics in their electronic ground
states. Next, desorption dynamics of the species involved is
considered and energy redistribution is analyzed. Finally, conclusions
are drawn.\\

\section{Methods}

\subsection{Intermolecular Interactions}
Reactive molecular dynamics simulations require potential energy
surfaces (PESs) that allow bond formation and bond
breaking.\cite{msarmd} Two such PESs were considered in the present
work. One was based on fitting Morse potentials $V(r) = D_e (1-\exp{
  \left[ -\beta(r-r_0)^2 \right]})$ to the N/C--O interaction based on
{\it ab initio} calculations using the MOLPRO\cite{molpro:2020} suite
of programs at the MRCI/aug-cc-pVTZ level of
theory.\cite{werner1988efficient,dunning1989gaussian} The bending
force constants were adapted to match with experimental
frequencies. The bending frequency for CO$_2$ and NO$_2$ are at 660
cm$^{-1}$ and 750 cm$^{-1}$ respectively which compares with 667
cm$^{-1}$ and 750 cm$^{-1}$ from experiments. For the angular
potential (the OCO/ONO bend), the parameters were those from the
CHARMM36 force field\cite{best2012optimization} and this PES is
referred to as MMH (for Morse-Morse-Harmonic). Such an approach is
similar to that used previously for oxygen-oxygen recombination on
amorphous solid water.\cite{MM.oxy:2019,MM.o2:2020} The second set of
PESs were reproducing kernel Hilbert Space (RKHS) representations of
reference energies computed using MOLPRO\cite{molpro:2020}. For CO$_2$
and NO$_2$, CCSD(T)-F12/aug-cc-pVTZ-F12\cite{MM.genesis:2021} and
MRCI+Q/aug-cc-pVTZ \cite{MM.NO2.2020} level ground state PESs were
used to run the dynamics.\\

\noindent
At the low temperatures prevalent in the interstellar medium diffusion
is a major driving force for chemical reactions. To provide an
accurate depiction of this phenomena, the fluctuating point charge
(FPC) model is used for CO, NO and atomic oxygen. In such a model the
point charges of CO, NO and atomic oxygen fluctuate as a function of
the OC/ON---O separation, see Figure S1. Mulliken
charges were obtained from DFT (M062x/aug-cc-pvtz) calculations of
snapshots from the MD simulations with OC/NO---O adsorbed on a cluster
containing the 10 nearest H$_2$O molecules with standard van der Waals
parameters from CHARMM36 force field.\cite{best2012optimization} After
inspection of the charge variations depending on
geometry, a sigmoidal function parametrization $y = a +
\frac{b-a}{1+e^{\frac{c-r}{d}}}$ was used to determine charges as a
function of $r$, where $r$ is the distance between ``C/N" of OC/ON
diatomics and atomic oxygen. For distances $r > 3.0$ \AA\/ the
diatomics and atomic oxygen behave as neutrals on the water surfaces
which significantly affects their diffusion behaviour.\\

\subsection{Molecular Dynamics Simulations}
\begin{figure}[h!]
\centering \includegraphics[scale=0.4]{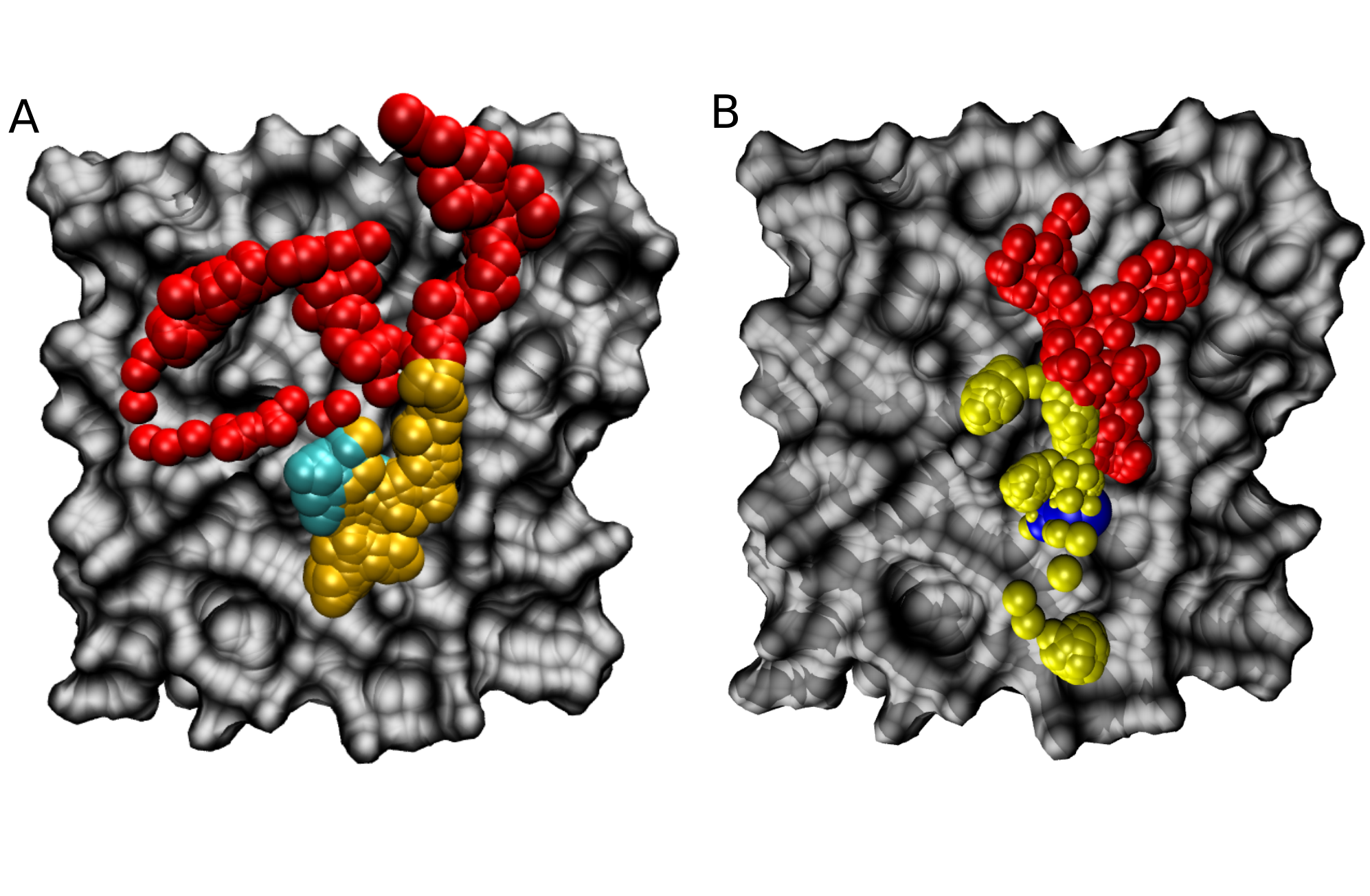}
 \caption{The simulation system for studying the CO$_2$ (Panel A) and
   NO$_2$ (Panel B) recombination reaction. Panel A color code: oxygen
   (red), CO (cyan), CO$_2$ (orange); Panel B color code: oxygen
   (red), NO (blue), NO$_2$ (yellow). Before and after recombination
   all species diffuse on the water surface in this case.}
 \label{fig:vmd}
\end{figure}

All molecular dynamics (MD) simulations were carried out using the
CHARMM suite of programs\cite{charmm.prog} with provisions for bond
forming reactions through MMH or RKHS
representations,\cite{MM.rkhs:2017} see above. The simulation system,
see Figure \ref{fig:vmd}, consisted of an equilibrated cubic box of
amorphous solid water containing 1000 water molecules with dimension
$31 \times 31 \times 31$ \AA\/$^3$. Simulations were started from an
existing, equilibrated ASW
structure\cite{MM.oxy:2018,MM.oxy:2019,MM.genesis:2021} by adding
adsorbates (CO, NO, O) on top of ASW.\\

\noindent
In the following, the coordinates are the diatomic stretch $r$, the
separation $R$ between the center of mass of the diatom and the oxygen
atom, and $\theta$ is the angle between $\vec{R}$ and $\vec{r}$.
Initial conditions (coordinates and velocities) were generated for a
grid of angles $\theta$ and separations $R$. With CO/NO and O
constrained at given values of $R$ and $\theta$, first 750 steps of
steepest descent and 100 steps Adopted Basis Newton-Raphson
minimization were carried out. This was followed by 50 ps of heating
dynamics to 50 K. Then, 100 ps equilibration with $R$ and $\theta$
still constrained dynamics simulations were carried out at 50 K. From
each of the runs coordinates and velocities were saved regularly to
obtain initial conditions for each combination of angle and
distance. Production simulations 500 ps in length were then run from
saved coordinates and velocities in the $NVE$ ensemble. Data
(energies, coordinates and velocities) were saved every 0.5 ps for
subsequent analysis.\\

\subsection{Analysis}
To characterize energy flow between the
activated internal modes of newly formed species and the surface water molecules,
the vibrational density of states (vDOS) was
analyzed.\cite{Gaigeot2007vdos,ferrero2023NH3vdos} For this, the
Fourier transform of the hydrogen atoms normalized velocity
autocorrelation function was determined \cite{futrelle1971vdos}
according to
    \begin{equation}
      I_{\rm vDOS}(\omega) = \int_{0}^{T_c} \frac{\langle v(0) \cdot
        v(t) \rangle}{\langle v(0) \cdot v(0) \rangle} e^{-i2 \pi
        \omega t} dt
  \end{equation}
where $v$ denotes the velocity vector of the hydrogen atom and $T_{\rm
  c} = 1$ ps.\\

\section{Results}

\subsection{Exploratory Simulations}
\begin{figure}[h!]
\centering \includegraphics[scale=0.42]{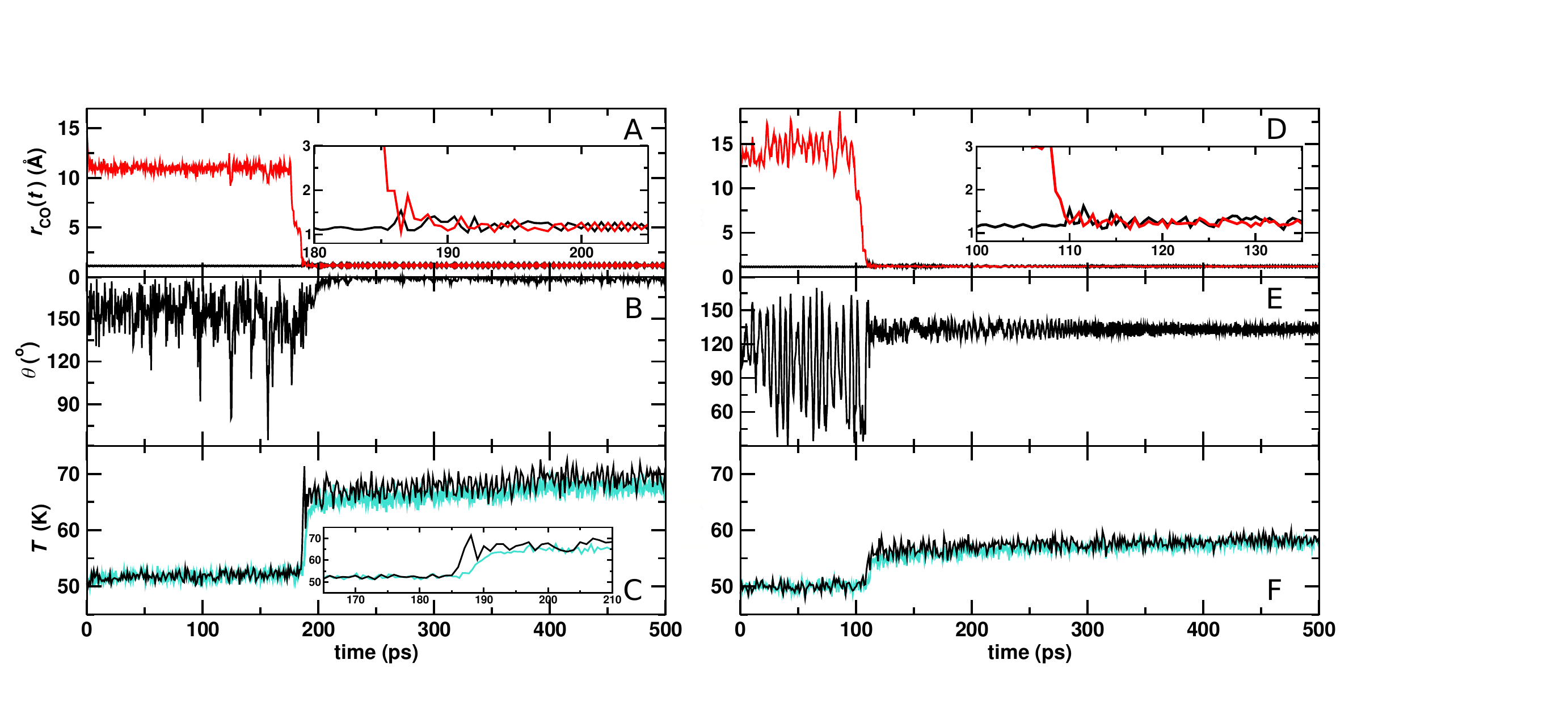}
\caption{Recombination of CO + O to form CO$_2$ (left column) and NO +
  O to NO$_2$ (right column) in ground state. Panel A and D: O$_{\rm
    B}$--C/N (red) and O$_{\rm A}$--C/N (black) separation; Panel B
  and E: O--C/N--O angle; Panel C and F: Temperature of the ASW (cyan)
  and the full system (black) before and after recombination. Upon
  recombination both O--C/N stretches exhibit equally pronounced
  excitation (shown in the insets).}
 \label{fig:timeseries}
\end{figure}

A representative 500 ps trajectory for CO$_{\rm A}$ + O$_{\rm B}$
recombination is shown in Figure \ref{fig:timeseries} (left
column). Initially, the C$\cdots$O$_{\rm B}$ separation is $\sim$ 12
\AA\/ (see panel A). For the first 190 ps, oxygen diffusion on the
water surface can be seen. Upon recombination at $t \sim 190$ ps the
CO stretch and the OCO (panel B) angle are highly excited and relax
during the following few picoseconds to average values around the
CO$_2$ equilibrium geometry. The CO$_2$ product remains in an
internally excited state for a considerably longer time scale, see
panel B. The temperatures of the ASW (cyan) and the full system
(black) are determined by the kinetic energies of the molecules. A
prominent peak in the temperature of full system is observed right
after recombination whereas the warming of cool water surface is
gradual (panel C inset).\\

\noindent
Figure \ref{fig:timeseries} (right column) shows a NO$_{\rm A}$ +
O$_{\rm B}$ recombination trajectory.  The ground state NO$_2$($^2$A')
has a nonlinear geometry with N-O distances of 1.2 \AA\/ and O-N-O
angle of $134^{\circ}$. In this case, recombination takes place after
$\sim 110$ ps and the amount of energy released is half compared to
the CO$_2$. This is due to the different stabilization energies of the
triatomics with respect to the CO/NO+O asymptote which are 7.71 eV for
CO$_2$ (for $\theta = 180^\circ$) and 3.24 eV for NO$_2$ (for $\theta
= 135^\circ$), respectively.\\

\subsection{Reaction Probabilities for CO$_2$ and NO$_2$ Formation}

\begin{figure}[h!]
\centering \includegraphics[scale=1.0]{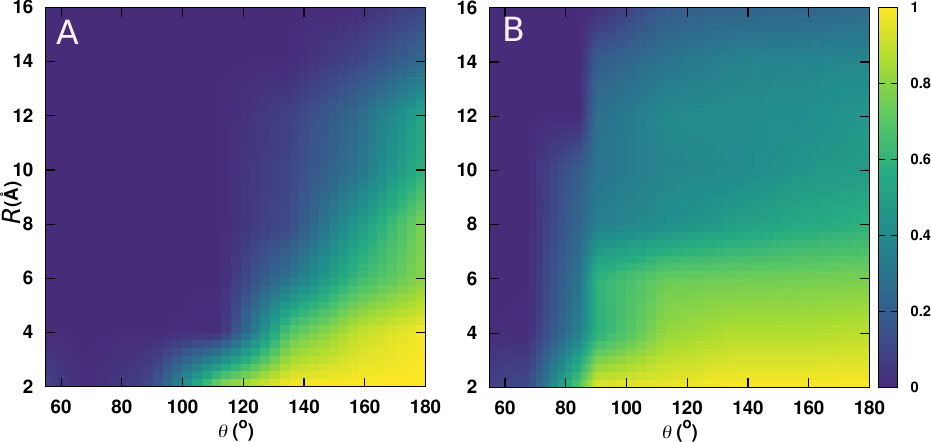}
 \caption{KDE density map for the average CO$_2$ formation probability
   depending on initial $(R,\theta)$ using the MMH (Panel A) and RKHS
   (Panel B) PES. The color palette indicates the probability of the
   reaction, with yellow representing $P = 1$ and blue representing $P
   = 0$. The COO formation probability using RKHS PES is shown in
   Figure S2.}
 \label{fig:heatmapco2}
\end{figure}

\begin{figure}[h!]
\centering \includegraphics[scale=1.0]{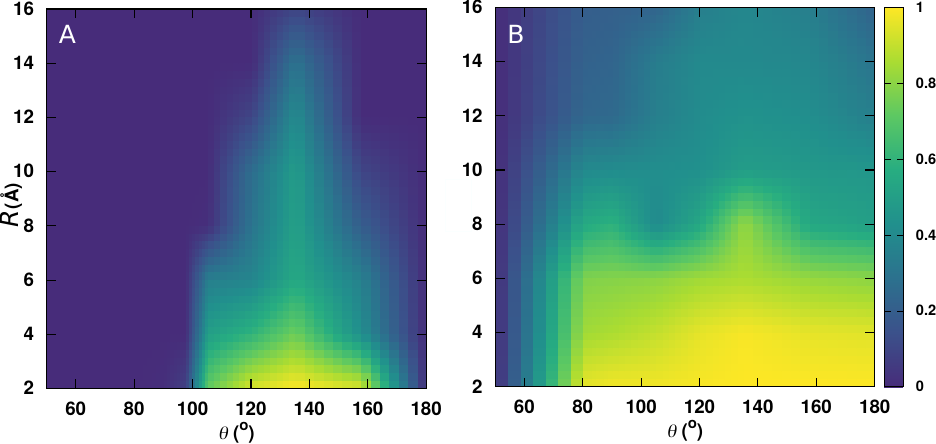}
 \caption{The NO$_2$ formation probability as a function of $R$ and
   $\theta$ using MMH (Panel A) and RKHS (Panel B) PES. The color
   palette indicates the probability of the reaction, with yellow
   representing $P = 1$ and blue representing $P = 0$.}
 \label{fig:heatmapno2}
\end{figure}

\noindent
The formation dynamics of CO$_2$ and NO$_2$ was followed from initial
conditions on a $(R,\theta)$ grid.  For each initial $R$ and $\theta$
configuration, 250 or 500 trajectories at 50 K were run for 500
ps. Depending on the initial $(R,\theta)$ values heat maps of the
formation probability were determined and a kernel density estimate
(KDE) was used for smoothing, see Figures \ref{fig:heatmapco2} and
\ref{fig:heatmapno2}.\\

\noindent
Figure \ref{fig:heatmapco2} reports probability heat maps for the
likelihood to form CO$_2$ depending on initial $(R, \theta)$ using
both MMH (Panel A) and RKHS (Panel B) PESs from simulations 500 ps in
length. With the MMH PES (Panel A) for $R \sim 2$ to 4.5 the
recombination probability is $P \sim 1$ near $180^{\circ}$ and
vanishes for $\theta \sim 90^{\circ}$. For $\theta = 180^\circ$, $P$
changes from unity for $R = 4$ to $P \sim 0$ for 14 \AA\/ on the 500
ps time scale. For $R> 4$ \AA\/ and $\theta \le 120^\circ$ the
formation probability is $P =0$.\\

\noindent
Using the RKHS PES, which is a considerably more accurate
representation of the O+CO interaction than the MMH parametrization,
the reaction probabilities depending on initial $(R, \theta)$ is
larger throughout, see Figure \ref{fig:heatmapco2} B. In particular
along the angular coordinate $\theta$ the probability for formation of
CO$_2$ is increased. These differences are due to the different
topographies of the two PESs. Also, the angular dependence is observed
for $R < 7$\AA\/ and for larger distances $R > 7$\AA\/ a plateau is
observed. This plateau implies that the reaction probability is
primarily governed by diffusion for large distances and does not
depend on the initial angular term on the 500 ps time scale.\\

\noindent
Due to the more realistic and improved charge model for the reactants,
their diffusivity increases. For simulations on the 500 ps this leads
to a nonvanishing reaction probability for initial separations of $R =
16$ \AA\/. This contrasts remarkably with earlier
simulations\citet{MM.genesis:2021} that maintained the charges on the
oxygen atom and CO molecule fixed at $\pm 0.3$e (corresponding to
their charges in CO$_2$) which suppressed the reactants' mobility and  reaction probability for
$R > 5$ \AA\/ for simulations on the same time scale. Overall, using
the more realistic charge models from the present work increases the
recombination probability of the species involved.\\

\noindent
The recombination probability for NO$_2$ formation is reported in
Figure \ref{fig:heatmapno2}. Panel A shows the 2-dimensional
recombination probability using the MMH PES. For NO$_2$ the minima
energy structure has $\theta = 134^\circ$ which leads to $P \sim 1$
for $\theta \in [125,145]^\circ$ and $R \leq 4$ \AA\/ which gradually
decreases to $P \sim 0$ as $R$ increases. Using the considerably more
accurate RKHS representation the recombination probability is larger
throughout, see Figure \ref{fig:heatmapno2}B. Formation probability $P
\sim 1$ is found for $\theta \in [90,180]^\circ$ and $R \leq 8$
\AA\/. These differences are due to the different topographies of the
two PESs. Similar to CO$_2$, the recombination probability is not
equal to zero at $R = 16$ \AA\/ on 500 ps time scale and a plateau is
observed for $R > 8$\AA\/.\\

\noindent
Although the structures of CO$_2$ and NO$_2$ differ (linear versus
bent), the shape of the underlying PES is reflected for both in
geometry-dependence of the rebinding probability for shorter
$R$. Whereas for larger distances, $P$ is purely governed by diffusion
and independent of $\theta$.\\

\noindent
The RKHS PES also supports the COO conformation and results in the
formation and stabilization of COO. This intermediate holds
significant interest as it can decay to C + O$_2$. The probability
heat map for the likelihood to form COO depending on $R$ and $\theta$
is shown in Figure S2. Finally, the atom exchange
reaction is found for 7 \% of all simulation using the MMH PES for
NO/CO + O. Atom exchange with RKHS is not observed for CO+O whereas
0.4 \% simulations show atom exchange for NO + O collision.\\

\subsection{Energy Dissipation upon XO$_2$ Relaxation} 
Both association reactions considered here are exothermic. Upon
recombination, products are formed in a highly excited internal state
(vibration and rotation) and their relaxation depends on the coupling
between excited stretching/bending modes of the newly formed species
and phonon and internal modes of the underlying water ice
surface.\citet{ferrero2023NH3vdos} To allow energy transfer between the
adsorbate and the internal and phonon modes of ASW, the
reparametrized\citet{Burnham97p6192,MM.ice:2008} flexible Kumagai,
Kawamura, and Yokokawa (KKY) water model\citet{kky_orig} was used in
these simulations. This model had been successfully used to study
vibrational relaxation of solvated cyanide\citet{MM.cn:2011} and the
relaxation of O$_2$ formed on ASW.\citet{MM.oxy:2019}\\

\noindent
To study energy dissipation following recombination several hundred
O+XO$\rightarrow$XO$_2$ trajectories were run for 500
ps. Subsequently, 280 recombination trajectories were analyzed for
CO$_2$ and NO$_2$, respectively. First, the normalized final kinetic
energy distribution $P(E_{\rm kin}^{\rm fin})$ at 450 ps after
recombination was determined for the products, see Figure
S3. The $P(E_{\rm kin}^{\rm fin})$ for CO$_2$ and NO$_2$
extend from low-$E_{\rm kin}^{\rm fin}$ ($\sim 1$ kcal/mol) to 60
kcal/mol. For NO$_2$ more than 80 \% of the products contain $E_{\rm
  kin}^{\rm fin} < 5$ kcal/mol after 450 ps whereas for CO$_2$ this
fraction is considerably smaller. In other words, relaxation of the
highly excited internal modes is more effective for NO$_2$ compared
with CO$_2$. Possible reasons for such rapid relaxation are the
shallow minimum for NO$_2$ compared to CO$_2$ and/or more efficient
coupling of the vibrational modes of NO$_2$ to the surrounding water
molecules.\\

\begin{figure}[h!]
    \centering
    \includegraphics[scale=0.31]{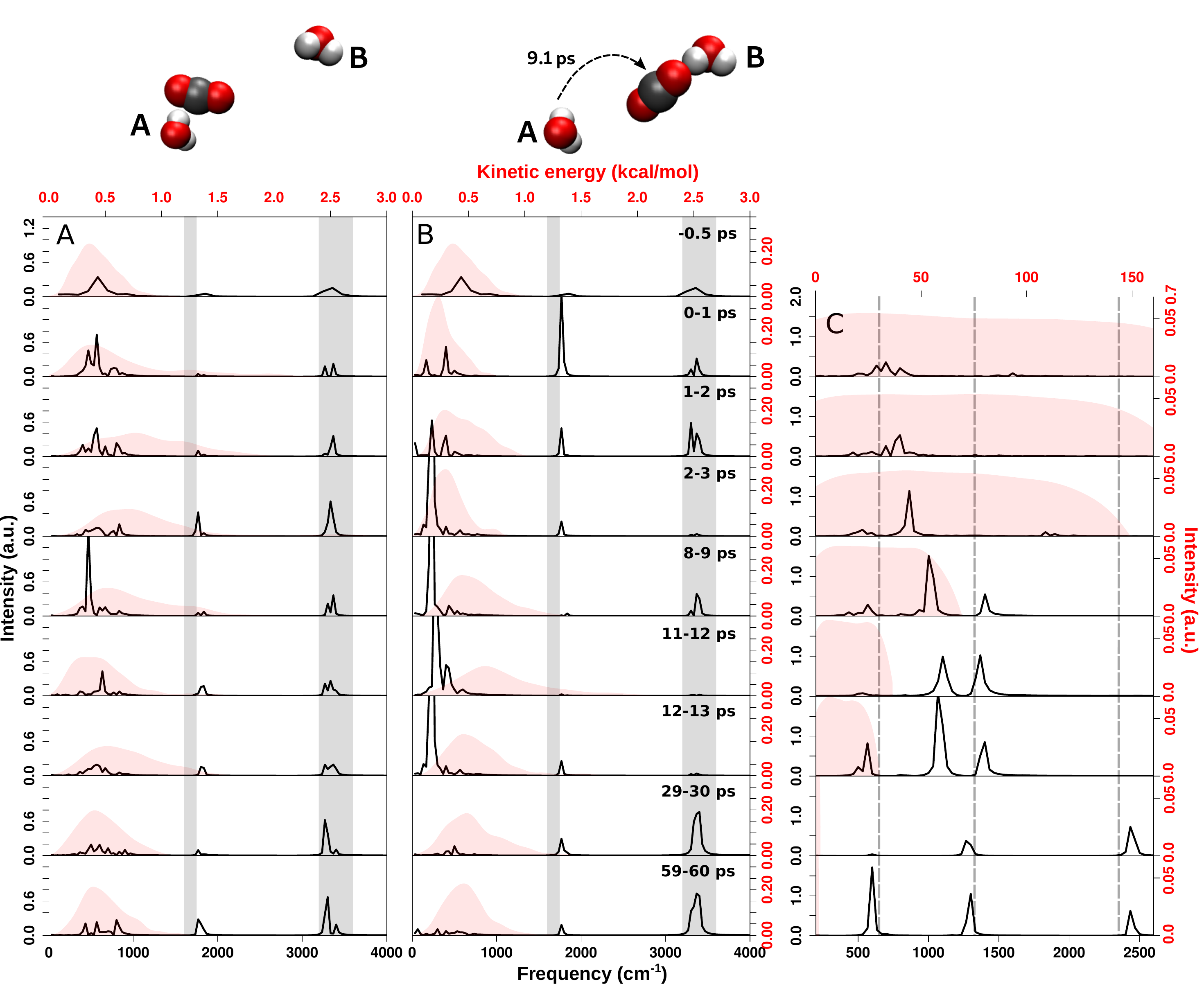}
    \caption{CO$_2$ recombination: The vibrational density of states
      power spectrum (black) and kinetic energy distribution (red) for
      water molecules A and B, W$_{\rm A}$ (panel A) and W$_{\rm B}$
      (panel B), on the ASW surface at delay times 1, 2, 3, 9, 12, 13,
      30, and 60 ps after formation of CO$_2$. Panel C reports the
      vDOS and kinetic energy distribution for CO$_2$. In panels A and
      B the shaded bands at $1600-1750$ and $3200-3600$ cm$^{-1}$
      correspond to bending and OH-stretching
      modes.\citet{hagen:1981,yu:2020,devlin:1990} Top panel: Averaged
      vDOS spectrum and kinetic energy distribution from 10 randomly
      chosen surface water molecules from data before recombination
      takes place. Subsequent panels are labeled with the time
      interval after CO$_2$ recombination.}
    \label{fig:energy-co2}
\end{figure}

\noindent
Next, for one CO$_2$-forming trajectory the kinetic energy
distribution and vDOS of two water molecules W$_{\rm A}$ and W$_{\rm
  B}$ and the CO$_2$ molecule were determined for windows 1 ps in
length. The time of recombination was set to zero. Water molecules
W$_{\rm A}$ and W$_{\rm B}$ were chosen according to the following
criterion: for the first 9 ps, W$_{\rm A}$ is the water molecule
nearest to the newly formed CO$_2$. At 9.1 ps, the CO$_2$ molecule
diffuses from one adsorption site on the ASW to a neighboring site
which results in W$_{\rm B}$ becoming the closest water molecule. As a
reference point for the analysis, the averaged $E_{\rm kin}$ and vDOS
for 10 water molecules located away from the recombination site was
determined for the 0.5 ps prior to CO$_2$-formation, see top row in
Figure \ref{fig:energy-co2}.\\

\noindent
For water molecule W$_{\rm A}$, the maximum of $P(E_{\rm kin})$ (red
shaded area) shifts to higher $E_{\rm kin}$ and its width increases
considerably for the first 9 ps, after which relaxation to the thermal
average is observed until 60 ps, see Figure \ref{fig:energy-co2}A. In
contrast, $P(E_{\rm kin})$ for W$_{\rm B}$ (Figure
\ref{fig:energy-co2}B) remains comparable to the thermal distribution
for the first 8 ps after which the position of the maximum and the
width increase. This is due to energy transfer between the partially
relaxed CO$_2$ molecule and W$_{\rm B}$ after diffusion of the
CO$_2$. After 13 ps and until 60 ps the position of the maximum
remains unchanged and the width of $P(E_{\rm kin})$ decreases such
that it reaches equilibrium again. As judged from the vDOS, the water
stretch, bend and libration modes for W$_{\rm A}$ and W$_{\rm B}$ are
populated for all time intervals. The two internal modes do not shift
in frequency relative to the thermal equilibrium and their frequency
is covered by the experimentally determined frequency
range.\citet{hagen:1981,yu:2020,devlin:1990}\\

\noindent
The CO$_2$ molecule formed is initially highly excited which leads to
a broad, unstructured $P(E_{\rm kin})$, top panel in Figure
\ref{fig:energy-co2}C. Over the next 9 ps the width of $P(E_{\rm
  kin})$ rapidly decreases due to energy exchange with W$_{\rm A}$ and
other surrounding water molecules and the phonon modes of the
ASW. Cooling of CO$_2$ continues after diffusion to the neighboring
site and within 60 ps after recombination the $P(E_{\rm kin})$ for
CO$_2$ is comparable to that of the surrounding water molecules. For
the vDOS a rather different picture than for water emerges. The
vibrational frequencies for CO$_2$ are at 667 (bend), 1333 (symmetric
stretch), 2349 (antisymmetric stretch) cm$^{-1}$ from
experiments\citet{shimanouchi:1978} (see dashed vertical lines) and
647, 1374, 2353 cm$^{-1}$ from calculations using the reactive PES, in
reasonably good agreement with experiment. At the end of the
observation window at 60 ps the three modes are clearly
visible. However, for times shortly after recombination only
frequencies below $\sim 1000$ cm$^{-1}$ appear in the
vDOS. Specifically, the antisymmetric stretch mode only manifests
itself after 19 ps (see Figure S4 for
detailed analysis) because the reactive PES for CO$_2$ is fully
anharmonic and for the first few ps after recombination the highly
excited stretching motions sample the narrowly spaced energies close
to dissociation.\\

\begin{figure}[h!]
    \centering
    \includegraphics[scale=0.3]{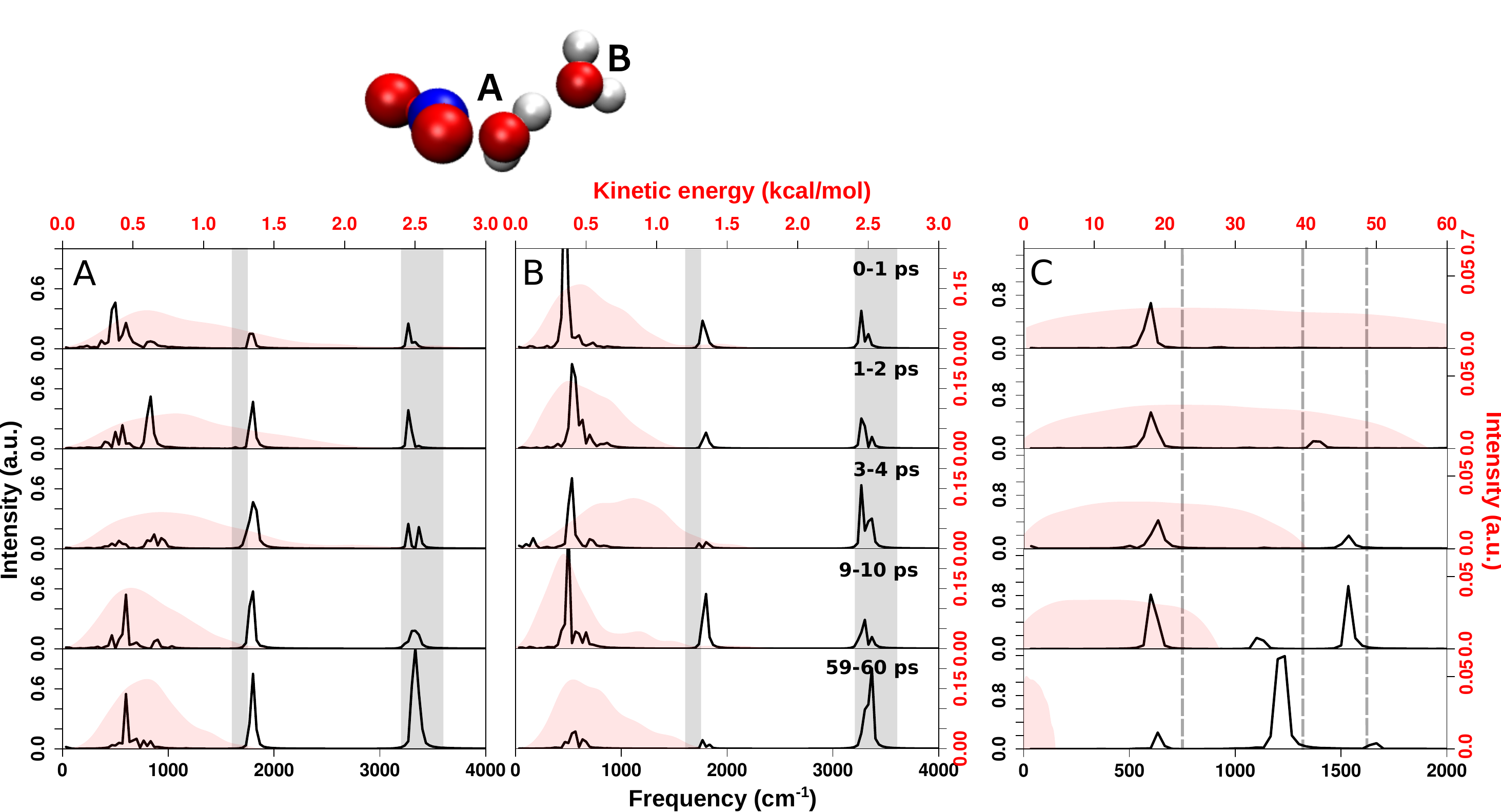}
    \caption{Vibrational density of states power spectrum (in black)
      and $P(E_{\rm kin})$ (red) for water molecules A and B, W$_{\rm
        A}$ (panel A) and W$_{\rm B}$ (panel B), on the ASW surface at
      delay times 1, 2, 4, 10, and 60 ps after the formation of
      NO$_2$. The IR bands at $1600-1750$ and $3200-3600$ cm$^{-1}$
      correspond to bending and OH-stretching modes.\citet{hagen:1981,
        yu:2020, devlin:1990} Panel C reports the kinetic energy
      distribution (red) and the vDOS of the NO$_2$ molecule.}
    \label{fig:energydiss-no2}
\end{figure}

\noindent
For studying energy transfer between neighboring water molecules on
the ASW, a slightly different approach was used for NO$_2$
recombination. For one NO$_2$-forming trajectory, the kinetic energy
distribution and vDOS of water molecules W$_{\rm A}$ and W$_{\rm B}$
and NO$_2$ were determined. Here, W$_{\rm A}$ represents the water
molecule closest to NO$_2$, while W$_{\rm B}$ is nearest to W$_{\rm
  A}$, see Figure \ref{fig:energydiss-no2}. For W$_{\rm A}$ the width
of $P(E_{\rm kin})$ is considerably broader than the thermal
distribution already for the first ps (0-1 ps), see top panel in
Figure \ref{fig:energydiss-no2}A, and narrow only at 9 ps after
recombination. In contrast, $P(E_{\rm kin})$ for W$_{\rm B}$ which is
not in direct contact with the recombining NO$_2$ is initially
comparable to thermal equilibrium but widens after 3 ps (Figure
\ref{fig:energydiss-no2}B). Within the next 6 ps the width decreases
again and equilibrium is established within 60 ps for both water
molecules. The peak maxima shift to higher $E_{\rm kin}$ at early
times and develop towards thermal equilibrium on the 60 ps time
scale. The kinetic energy distribution for NO$_2$ starts out broad (60
kcal/mol) and transfers more than half of this energy to the
environment within 10 ps of recombination (Figure
\ref{fig:energydiss-no2}C). Within 60 ps $P(E_{\rm kin})$ is close to
thermal equilibrium.\\

\noindent
The experimentally measured\citet{arakawa:1958_NO2-IR} vibrational
frequencies for gas phase NO$_2$ are at 750 (bend), 1318 (symmetric
stretch), 1618 (antisymmetric stretch) cm$^{-1}$. A blue shift over
time is observed as the NO$_2$ molecule loses energy. After 60 ps, it
begins to probe the low-energy regions of the potential, revealing
distinct peaks. For NO$_2$, the final $E_{\rm kin}$ fluctuates around
$\sim 5$ kcal/mol, causing the peak to shift from its equilibrium
features. During the initial picoseconds, there is a notable
population in the region of the water libration modes compared to
bending and stretching. However beyond 60 ps, as the system approaches
equilibrium, the range of the libration modes is less
populated/unpopulated. In contrast, the stretching modes gain
intensity.\\

\noindent
Comparing specifically the time evolution of $P(E_{\rm kin})$ for
CO$_2$ and NO$_2$ it is observed that the two products differ in the
efficiency and speed of cooling. For NO$_2$ the fraction of products
formed with $E_{\rm kin} \sim 5$ kcal/mol within $\sim 500$ ps is
twice as large compared with CO$_2$. This may be related to the lower
energy of formation (3.24 eV vs. 7.71 eV) but also due to the overall
lower vibrational energies of the fundamentals for NO$_2$ compared
with CO$_2$.\\

\begin{figure}[h!]
\centering
   \includegraphics[scale=0.4]{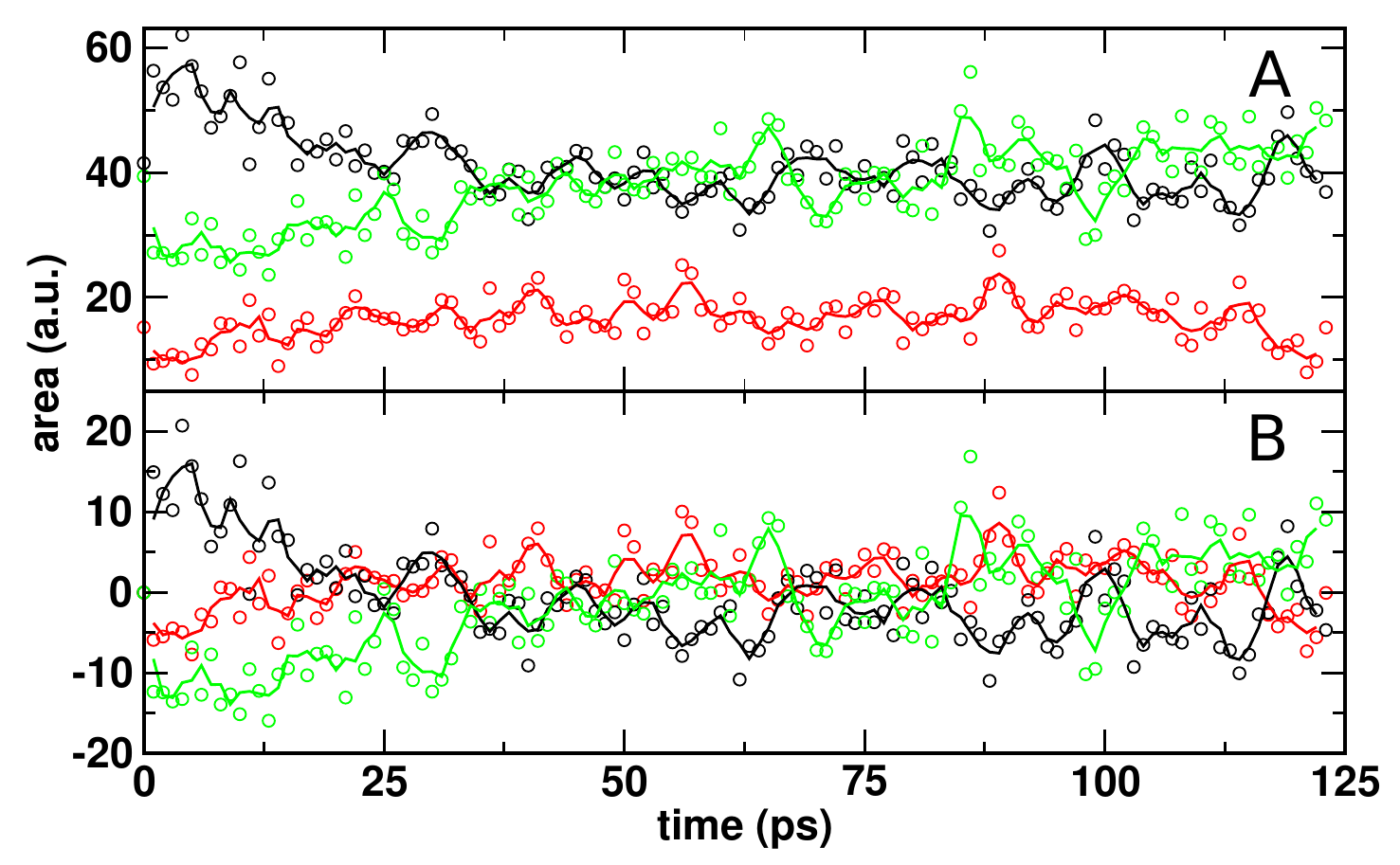}
    \caption{Time evolution of the integrated change of
      vDOS curves averaged over 100 water molecules for libration
      (black), bending (red), and stretching modes (green) of
      water. In Panel A, the equilibrium, thermally averaged area
      before recombination is the reference for
      calculating the difference at time $t$ whereas for Panel B the peak area right after 
      excitation ($t = 1$ps) is the reference. Here, solid lines
      represent the 4-point moving average.}
    \label{fig:peakarea-co2}
\end{figure}

\noindent
For a more mode-specific analysis of energy transfer between the
formed CO$_2$ and the surrounding water molecules the time evolution
of the integrated vDOS in three frequency ranges was considered. These
frequency ranges correspond to water libration ($0-1000$ cm$^{-1}$),
bending ($1500-2000$ cm$^{-1}$) and stretching ($2800-3800$
cm$^{-1}$), in Figures \ref{fig:energy-co2} and
\ref{fig:energydiss-no2}. For analysis, water molecules within 10
\AA\/ in the $x-$direction and 5 \AA\/ in the $y/z-$ direction located
around the recombination site were used, see Figure
S5. Then the vibrational density of states (vDOS)
was computed for each water molecule separately, and the resulting
peak areas were averaged over 100 water molecules. Figure
\ref{fig:peakarea-co2}A reports the difference between the equilibrium
averaged peak area and the area at time $t$ after recombination.  For
better comparison, the difference between peak area right after
excitation and the area at time $t$ is shown in Panel B. The
integrated intensity of the low-frequency libration mode (black line)
initially increases, followed by a decrease after 12 ps, and
eventually reaching a plateau. Meanwhile, the intensity of the
stretching modes (green) displays a progressive rise over the first 50
ps, followed by fluctuations around a specific value. In comparison,
the intensity of bending modes remains comparable throughout the
observed time frame (similar observation in Figure
S6). The time series show that the libration
mode first acquires and stores energy and subsequently releases it,
whereas the stretching modes pick up energy on the 100 ps time scale
and retain it beyond that. This was also observed for NH$_3$ formation
on water ice.\citet{ferrero2023NH3vdos}\\

\subsection{Species Desorption on ASW}
To explicitly probe desorption, velocities for the adsorbants (CO, NO,
CO$_2$, and NO$_2$) were drawn from a Maxwell-Boltzmann distribution
and scaled along the $x-$direction, perpendicular to the water
surface.\citet{MM.o2:2020} For the water molecules the velocities were
those from an equilibrium simulation at 50 K and the dynamics of the
system was followed for 100 ps. If within this time scale the
adsorbant remained on the ASW, it was considered physisorbed. However,
for sufficiently large scaling of the velocity vector the adsorbant
leaves the ASW from which the desorption energy can be
estimated.\citet{MM.o2:2020} Average desorption energies were
determined from initiating the dynamics for different initial
positions of the adsorbant on the ASW.\\

\noindent
Computed desorption energies ($E_{\rm des}$) for CO and NO from the
ASW surface were $3.5 \pm 0.7$ kcal/mol (1812 K or 156 meV) and $3.1
\pm 0.5$ kcal/mol (1534 K or 132 meV), respectively. Earlier MD
simulations reported\cite{MM.genesis:2021} CO-desorption energies
between 3.1 and 4.0 kcal/mol (1560 K to 2012 K or 130 meV to 170 meV),
compared with 120 meV (2.8 kcal/mol or 1392 K) from
experiments.\citet{cuppen:2013} It was also found that the desorption
energy of CO from ASW depends on CO coverage with ranges from $E_{\rm
  des} = 1700$ K for low to $E_{\rm des} = 1000$ K for high coverage,
respectively,\citet{he:2016} which supports the simulation
results.\cite{MM.genesis:2021} On non-porous and crystalline water
surfaces submonolayer desorption energies for CO are 1307 K and 1330 K
($\sim 115$ meV), respectively.\citet{noble:2012} For NO, experiments
found a desorption energy of 1300 K (2.6 kcal/mol or 112 meV) using
temperature programmed desorption
measurements\citet{minissale2019experimental} after exposure of
NO/H$_2$O ice to atomic oxygen. Thus, for both diatomics the present
findings are consistent with experiments and earlier simulations.\\

\noindent
For the triatomics, average desorption energies for CO$_2$ and NO$_2$
from ASW were $9.0 \pm 1.9$ kcal/mol (4529 K or 390 meV) and $13.1 \pm
3.4$ kcal/mol (6592 K and 568 meV). For CO$_2$ one experiment reported
$E_{\rm des} = 2490 \pm 240$ K \citet{galvez2007co2desorption} ($4.9
\pm 0.47$ kcal/mol or $214 \pm 20$ meV) and a compilation of
literature data reports values between 4.5 and 6.0
kcal/mol.\citet{wakelam.binding:2017} On the other hand, for NO$_2$ no
experimentally measured data is available;\citet{wakelam.binding:2017}
however, the desorption temperatures of NO$_2$ and H$_2$O from the
same ice surface were found to be comparable which indicates that
NO$_2$ interacts strongly with the surface.\citet{linnartz:2014-2}
Overall, for all four species relevant to the present work the
computed $E_{\rm des}$ compare reasonably (CO$_2$) to favourably (CO,
NO) with reference values from experiments and previous calculations,
particularly in light of the fact that appreciable variations in
experimentally reported values exist.\citet{wakelam.binding:2017}\\

\subsection{Species Diffusion on ASW}
\begin{figure}[h!]
\centering 
\includegraphics[scale=0.80]{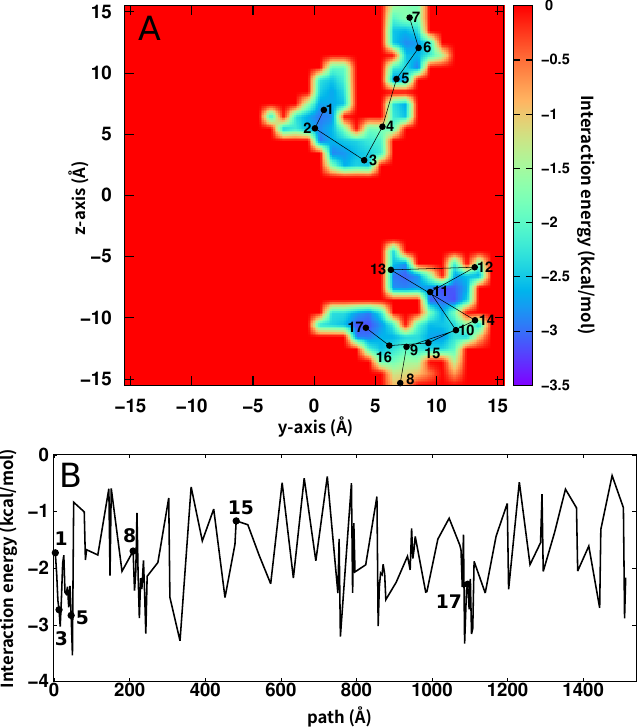}
\caption{Diffusion of CO on ASW. Panel A: 2D projection of CO
  interaction energy and diffusion path onto the ASW $y-z$ plane from
  a 30 ns long simulation at 50 K. Panel B: 1D projection of the path
  followed with some of the minima visited in panel A labeled with
  corresponding numbers. Between minima 15 and 17 the adsorbant
  repeatedly visits minima 8 to 17. The average diffusion barrier
  height is 1.15 kcal/mol compared with a range from 0.25 kcal/mol to
  1.85 kcal/mol from experiments.\citet{acharyya:2022,cuppen:2014}}
\label{fig:diffusion-co}
\end{figure}

To determine typical diffusion barrier heights, a 30 ns long
simulation of molecules (CO, NO, CO$_2$, and NO$_2$) on water surface
was run with velocities from a Maxwell-Boltzmann distribution at 50
K. For each frame, their interaction energy with the ASW surface was
determined and mapped onto the ASW surface to give a 2-dimensional
map, see Figures \ref{fig:diffusion-co} and
S7. Figure \ref{fig:diffusion-co}A shows the 2D
projection of CO interaction energy with ASW from a 30 ns long
simulation. The black solid line traces the path followed by the CO
molecule on top of the ASW surface. The total path length with several
minima enumerated and the interaction energy between CO and the ASW is
reported in Figure \ref{fig:diffusion-co}B. The diffusion barrier
heights vary between minima and are a consequence of ASW surface
roughness.\citet{MM.oxy:2018} On average, the activation barrier for
diffusion between neighboring minima is 1.15 kcal/mol. Experimentally,
a range of diffusion barriers was found to extend from $E_{\rm b} \sim
120 \pm 180$ K to $490 \pm 12$ K.\citet{kouchi:2020,acharyya:2022}
Another study reported $E_{\rm b}$ separately for weakly and strongly
bound sites to be $\sim 350$ K and $\sim 930$ K,
respectively.\citet{cuppen:2014} Hence, experimentally the diffusion
barriers range from 0.25 kcal/mol to 1.85 kcal/mol. The present
simulations support a distribution of barriers and are consistent with
the experimentally reported values. \\

\noindent
The calculated average activation barrier for diffusion of NO
(Figure S7A) between neighboring minima is 0.91
kcal/mol. From equilibrium trajectories at 50 K no diffusion for
CO$_2$ and NO$_2$ on the 75 ns time scale is observed.  For CO$_2$ one
experiment reported an average diffusion barrier of 2150 K (4.3
kcal/mol).\citet{he:2018} As a comparison, for CO the average
diffusion barrier height is $\sim 1$ kcal/mol which leads to a
transition time between neighboring minima of $\sim 1$ ns. Within an
Arrhenius picture for surface diffusion - which may not be entirely
appropriate at sufficiently low temperatures\citet{MM.oxy:2018} - a
barrier of $\sim 4$ kcal/mol corresponds to a diffusion time scale on
the order of several hundred ns. This is consistent with the present
findings that on the sub-100 ns time scale CO$_2$ does not diffuse on
ASW. For comparison, the diffusion of CO$_2$ and NO$_2$ on the water
surface after recombination is shown in Figure
S8.\\

\noindent
Earlier work already demonstrated that oxygen diffusion coefficients
are consistent with the experiment. Atomic oxygen\citet{MM.oxy:2018} on ASW
experiences diffusional barriers between $E_{\rm dif} = 0.2$ kcal/mol
and 2 kcal/mol (100 K to 1000 K) compared with an averaged value of
$E_{\rm dif} = 990_{\rm -360}^{\rm +530}$ K determined from
experiments.\citet{dulieu:2016}\\
  
\section{Discussion and Conclusion}
The present work underlines that larger molecules can form and
stabilize from atomic and molecular constituents on ASW. The
rate-limiting step for CO$_2-$ and NO$_2-$generation is the diffusion
of the reactant species. After recombination, both products
efficiently cool on the 10 picosecond time scale although a smaller fraction still remains in an internally hot state. Use of a fluctuating
charge model clarifies that on the nanosecond time scale recombination
can occur from atom--molecule separations up to $\sim 20$ \AA\/. Given
the much longer time scales available in interstellar chemistry the
areas covered by the reactants will increase in proportion. The energy
liberated upon recombination is efficiently transferred into
water-internal modes and lattice vibrations (phonons) on the
picosecond time scale. Of particular note is the vanishingly small
equilibrium diffusivity of CO$_2$ and NO$_2$ at low temperatures (here 50 K) which is consistent with
experiment, whereas immediately after recombination the triatomics diffuse
and roam wider parts of the ASW. For NO$_2$, 11 \% of
the recombined trajectories show desorption in 1 ns long simulations, whereas for CO$_2$, no
desorption is observed.\\

\noindent
Astrophysical implications of the present work include physical and
chemical aspects alike. Recombination reactions liberate energy to drive restructuring of the underlying surface. As a
consequence of recombination local heating of the ASW can take place
which also potentially increases the local diffusivity and leads to desorption
of reactive species
such as H or OH. This is particularly relevant as the cold average
temperatures of $\sim 10$ K restrict motion to the lightest particles
or driven by tunneling only.\citet{cuppen:2017_nonthermaldiffusion}
Because the surface morphology in part determines its chemical
activity, reconfiguration of the solid support is an important
determinant for downstream reactivity. The simulations also show that
following recombination, products undergo nonthermal surface diffusion
(see Figure S8) which leads to mass
transport across the surface to initiate chemistry at other
locations. This is particularly relevant because products can remain
in an internally excited state that can lead to breakup and radical
formation.\\

\noindent
In summary, the present work establishes that triatomics can be formed
and stabilized following atom+diatom recombination reactions. The
kernel-based potential energy surfaces combined with physically
meaningful electrostatic models allow to run statistically significant
numbers of trajectories and provide molecular-level insight into
energy relaxation and redistribution. This work provides a basis for
broader reaction exploration on ASW, in particular in light of recent
progress to represent high-dimensional, reactive PESs based on neural
networks.\cite{MM.criegee:2023}

\section*{Acknowledgments}
The authors gratefully acknowledge financial support from the Swiss
National Science Foundation through grant $200020\_188724$ and to the
NCCR-MUST.\\

\bibliography{astro3}

\providecommand{\latin}[1]{#1}
\makeatletter
\providecommand{\doi}
  {\begingroup\let\do\@makeother\dospecials
  \catcode`\{=1 \catcode`\}=2 \doi@aux}
\providecommand{\doi@aux}[1]{\endgroup\texttt{#1}}
\makeatother
\providecommand*\mcitethebibliography{\thebibliography}
\csname @ifundefined\endcsname{endmcitethebibliography}
  {\let\endmcitethebibliography\endthebibliography}{}
\begin{mcitethebibliography}{79}
\providecommand*\natexlab[1]{#1}
\providecommand*\mciteSetBstSublistMode[1]{}
\providecommand*\mciteSetBstMaxWidthForm[2]{}
\providecommand*\mciteBstWouldAddEndPuncttrue
  {\def\EndOfBibitem{\unskip.}}
\providecommand*\mciteBstWouldAddEndPunctfalse
  {\let\EndOfBibitem\relax}
\providecommand*\mciteSetBstMidEndSepPunct[3]{}
\providecommand*\mciteSetBstSublistLabelBeginEnd[3]{}
\providecommand*\EndOfBibitem{}
\mciteSetBstSublistMode{f}
\mciteSetBstMaxWidthForm{subitem}{(\alph{mcitesubitemcount})}
\mciteSetBstSublistLabelBeginEnd
  {\mcitemaxwidthsubitemform\space}
  {\relax}
  {\relax}

\bibitem[Herbst(2001)]{herbst:2001}
Herbst,~E. The chemistry of interstellar space. \emph{Chem. Soc. Rev.}
  \textbf{2001}, \emph{30}, 168--176\relax
\mciteBstWouldAddEndPuncttrue
\mciteSetBstMidEndSepPunct{\mcitedefaultmidpunct}
{\mcitedefaultendpunct}{\mcitedefaultseppunct}\relax
\EndOfBibitem
\bibitem[Wakelam \latin{et~al.}(2017)Wakelam, Bron, Cazaux, Dulieu, Gry,
  Guillard, Habart, Hornek{\ae}r, Morisset, Nyman, \latin{et~al.}
  others]{wakelam:2017}
Wakelam,~V.; Bron,~E.; Cazaux,~S.; Dulieu,~F.; Gry,~C.; Guillard,~P.;
  Habart,~E.; Hornek{\ae}r,~L.; Morisset,~S.; Nyman,~G., \latin{et~al.}  H$_2$
  formation on interstellar dust grains: The viewpoints of theory, experiments,
  models and observations. \emph{Mol. Astrophys.} \textbf{2017}, \emph{9},
  1--36\relax
\mciteBstWouldAddEndPuncttrue
\mciteSetBstMidEndSepPunct{\mcitedefaultmidpunct}
{\mcitedefaultendpunct}{\mcitedefaultseppunct}\relax
\EndOfBibitem
\bibitem[Hagen \latin{et~al.}(1981)Hagen, Tielens, and Greenberg]{hagen:1981}
Hagen,~W.; Tielens,~A.; Greenberg,~J. The infrared spectra of amorphous solid
  water and ice Ic between 10 and 140 K. \emph{Chem. Phys.} \textbf{1981},
  \emph{56}, 367--379\relax
\mciteBstWouldAddEndPuncttrue
\mciteSetBstMidEndSepPunct{\mcitedefaultmidpunct}
{\mcitedefaultendpunct}{\mcitedefaultseppunct}\relax
\EndOfBibitem
\bibitem[Jenniskens and Blake(1994)Jenniskens, and Blake]{jenniskens:1994}
Jenniskens,~P.; Blake,~D. Structural transitions in amorphous water ice and
  astrophysical implications. \emph{Science} \textbf{1994}, \emph{265},
  753--756\relax
\mciteBstWouldAddEndPuncttrue
\mciteSetBstMidEndSepPunct{\mcitedefaultmidpunct}
{\mcitedefaultendpunct}{\mcitedefaultseppunct}\relax
\EndOfBibitem
\bibitem[Bossa \latin{et~al.}(2012)Bossa, Isokoski, de~Valois, and
  Linnartz]{linnartz:2012}
Bossa,~J.~B.; Isokoski,~K.; de~Valois,~M.~S.; Linnartz,~H. Thermal collapse of
  porous interstellar ice. \emph{Astron. Astrophys.} \textbf{2012}, \emph{545},
  A82\relax
\mciteBstWouldAddEndPuncttrue
\mciteSetBstMidEndSepPunct{\mcitedefaultmidpunct}
{\mcitedefaultendpunct}{\mcitedefaultseppunct}\relax
\EndOfBibitem
\bibitem[Bar-Nun \latin{et~al.}(1987)Bar-Nun, Dror, Kochavi, and
  Laufer]{barnun:1987}
Bar-Nun,~A.; Dror,~J.; Kochavi,~E.; Laufer,~D. Amorphous water ice and its
  ability to trap gases. \emph{Phys. Rev. B} \textbf{1987}, \emph{35},
  2427--2435\relax
\mciteBstWouldAddEndPuncttrue
\mciteSetBstMidEndSepPunct{\mcitedefaultmidpunct}
{\mcitedefaultendpunct}{\mcitedefaultseppunct}\relax
\EndOfBibitem
\bibitem[He \latin{et~al.}(2016)He, Acharyya, and Vidali]{he:2016}
He,~J.; Acharyya,~K.; Vidali,~G. Binding Energy of Molecules on Water Ice:
  Laboratory Measurements and Modeling. \emph{Astrophys. J.} \textbf{2016},
  \emph{825}, 89\relax
\mciteBstWouldAddEndPuncttrue
\mciteSetBstMidEndSepPunct{\mcitedefaultmidpunct}
{\mcitedefaultendpunct}{\mcitedefaultseppunct}\relax
\EndOfBibitem
\bibitem[Kouchi \latin{et~al.}(2020)Kouchi, Furuya, Hama, Chigai, Kozasa, and
  Watanabe]{kouchi:2020}
Kouchi,~A.; Furuya,~K.; Hama,~T.; Chigai,~T.; Kozasa,~T.; Watanabe,~N. Direct
  Measurements of Activation Energies for Surface Diffusion of CO and CO$_2$ on
  Amorphous Solid Water Using In Situ Transmission Electron Microscopy.
  \emph{Astrophys. J. Lett.} \textbf{2020}, \emph{891}, L22\relax
\mciteBstWouldAddEndPuncttrue
\mciteSetBstMidEndSepPunct{\mcitedefaultmidpunct}
{\mcitedefaultendpunct}{\mcitedefaultseppunct}\relax
\EndOfBibitem
\bibitem[Oba \latin{et~al.}(2009)Oba, Miyauchi, Hidaka, Chigai, Watanabe, and
  Kouchi]{Oba09p464}
Oba,~Y.; Miyauchi,~N.; Hidaka,~H.; Chigai,~T.; Watanabe,~N.; Kouchi,~A.
  Formation of Compact Amorphous {H}$_2${O} Ice by Codeposition of Hydrogen
  Atoms with Oxygen Molecules on Grain Surfaces. \emph{Astrophys. J.}
  \textbf{2009}, \emph{701}, 464--470\relax
\mciteBstWouldAddEndPuncttrue
\mciteSetBstMidEndSepPunct{\mcitedefaultmidpunct}
{\mcitedefaultendpunct}{\mcitedefaultseppunct}\relax
\EndOfBibitem
\bibitem[Keane \latin{et~al.}(2001)Keane, Tielens, Boogert, Schutte, and
  Whittet]{keane:2001}
Keane,~J.; Tielens,~A.; Boogert,~A.; Schutte,~W.; Whittet,~D. Ice absorption
  features in the 5-8 $\mu$m region toward embedded protostars. \emph{Astron.
  Astrophys.} \textbf{2001}, \emph{376}, 254--270\relax
\mciteBstWouldAddEndPuncttrue
\mciteSetBstMidEndSepPunct{\mcitedefaultmidpunct}
{\mcitedefaultendpunct}{\mcitedefaultseppunct}\relax
\EndOfBibitem
\bibitem[Kouchi \latin{et~al.}(2021)Kouchi, Tsuge, Hama, Oba, Okuzumi, Sirono,
  Momose, Nakatani, Furuya, Shimonishi, \latin{et~al.} others]{kouchi:2021}
Kouchi,~A.; Tsuge,~M.; Hama,~T.; Oba,~Y.; Okuzumi,~S.; Sirono,~S.-i.;
  Momose,~M.; Nakatani,~N.; Furuya,~K.; Shimonishi,~T., \latin{et~al.}
  Transmission Electron Microscopy Study of the Morphology of Ices Composed of
  H$_2$O, CO$_2$, and CO on Refractory Grains. \emph{Astrophys. J.}
  \textbf{2021}, \emph{918}, 45\relax
\mciteBstWouldAddEndPuncttrue
\mciteSetBstMidEndSepPunct{\mcitedefaultmidpunct}
{\mcitedefaultendpunct}{\mcitedefaultseppunct}\relax
\EndOfBibitem
\bibitem[{Bossa, J.-B.} \latin{et~al.}(2014){Bossa, J.-B.}, {Isokoski, K.},
  {Paardekooper, D. M.}, {Bonnin, M.}, {van der Linden, E. P.}, {Triemstra,
  T.}, {Cazaux, S.}, {Tielens, A. G. G. M.}, and {Linnartz, H.}]{bossa:2014}
{Bossa, J.-B.},; {Isokoski, K.},; {Paardekooper, D. M.},; {Bonnin, M.},; {van
  der Linden, E. P.},; {Triemstra, T.},; {Cazaux, S.},; {Tielens, A. G. G.
  M.},; {Linnartz, H.}, Porosity measurements of interstellar ice mixtures
  using optical laser interference and extended effective medium
  approximations. \emph{Astron. Astrophys.} \textbf{2014}, \emph{561},
  A136\relax
\mciteBstWouldAddEndPuncttrue
\mciteSetBstMidEndSepPunct{\mcitedefaultmidpunct}
{\mcitedefaultendpunct}{\mcitedefaultseppunct}\relax
\EndOfBibitem
\bibitem[Bossa \latin{et~al.}(2015)Bossa, Mat{\'{e}}, Fransen, Cazaux, Pilling,
  Rocha, Ortigoso, and Linnartz]{bossa:2015}
Bossa,~J.-B.; Mat{\'{e}},~B.; Fransen,~C.; Cazaux,~S.; Pilling,~S.; Rocha,~W.
  R.~M.; Ortigoso,~J.; Linnartz,~H. Porosity and band-strength measurements of
  multi-phase composite ices. \emph{Astrophys. J.} \textbf{2015}, \emph{814},
  47\relax
\mciteBstWouldAddEndPuncttrue
\mciteSetBstMidEndSepPunct{\mcitedefaultmidpunct}
{\mcitedefaultendpunct}{\mcitedefaultseppunct}\relax
\EndOfBibitem
\bibitem[{Cazaux, S.} \latin{et~al.}(2015){Cazaux, S.}, {Bossa, J.-B.},
  {Linnartz, H.}, and {Tielens, A. G. G. M.}]{cazauxs:2015}
{Cazaux, S.},; {Bossa, J.-B.},; {Linnartz, H.},; {Tielens, A. G. G. M.}, Pore
  evolution in interstellar ice analogues - Simulating the effects of
  temperature increase. \emph{Astron. Astrophys.} \textbf{2015}, \emph{573},
  A16\relax
\mciteBstWouldAddEndPuncttrue
\mciteSetBstMidEndSepPunct{\mcitedefaultmidpunct}
{\mcitedefaultendpunct}{\mcitedefaultseppunct}\relax
\EndOfBibitem
\bibitem[Ioppolo \latin{et~al.}(2011)Ioppolo, Cuppen, and
  Linnartz]{Ioppolo:2011}
Ioppolo,~S.; Cuppen,~H.~M.; Linnartz,~H. Surface formation routes of
  interstellar molecules: hydrogenation reactions in simple ices. \emph{Rend.
  Lincei.} \textbf{2011}, \emph{22}, 211--224\relax
\mciteBstWouldAddEndPuncttrue
\mciteSetBstMidEndSepPunct{\mcitedefaultmidpunct}
{\mcitedefaultendpunct}{\mcitedefaultseppunct}\relax
\EndOfBibitem
\bibitem[Romanzin \latin{et~al.}(2011)Romanzin, Ioppolo, Cuppen, van Dishoeck,
  and Linnartz]{romanzin:2011}
Romanzin,~C.; Ioppolo,~S.; Cuppen,~H.~M.; van Dishoeck,~E.~F.; Linnartz,~H.
  Water formation by surface O$_3$ hydrogenation. \emph{J. Chem. Phys.}
  \textbf{2011}, \emph{134}, 084504\relax
\mciteBstWouldAddEndPuncttrue
\mciteSetBstMidEndSepPunct{\mcitedefaultmidpunct}
{\mcitedefaultendpunct}{\mcitedefaultseppunct}\relax
\EndOfBibitem
\bibitem[Chaabouni \latin{et~al.}(2012)Chaabouni, Minissale, Manicò, Congiu,
  Noble, Baouche, Accolla, Lemaire, Pirronello, and Dulieu]{Chaabouni:2012}
Chaabouni,~H.; Minissale,~M.; Manicò,~G.; Congiu,~E.; Noble,~J.~A.;
  Baouche,~S.; Accolla,~M.; Lemaire,~J.~L.; Pirronello,~V.; Dulieu,~F. Water
  formation through O$_2$ + D pathway on cold silicate and amorphous water ice
  surfaces of interstellar interest. \emph{J. Chem. Phys.} \textbf{2012},
  \emph{137}, 234706\relax
\mciteBstWouldAddEndPuncttrue
\mciteSetBstMidEndSepPunct{\mcitedefaultmidpunct}
{\mcitedefaultendpunct}{\mcitedefaultseppunct}\relax
\EndOfBibitem
\bibitem[Minissale \latin{et~al.}(2013)Minissale, Congiu, Baouche, Chaabouni,
  Moudens, Dulieu, Accolla, Cazaux, Manico, and
  Pirronello]{oxy.diff.minissale:2013}
Minissale,~M.; Congiu,~E.; Baouche,~S.; Chaabouni,~H.; Moudens,~A.; Dulieu,~F.;
  Accolla,~M.; Cazaux,~S.; Manico,~G.; Pirronello,~V. Quantum Tunneling of
  Oxygen Atoms on Very Cold Surfaces. \emph{Phys. Rev. Lett.} \textbf{2013},
  \emph{111}, 053201\relax
\mciteBstWouldAddEndPuncttrue
\mciteSetBstMidEndSepPunct{\mcitedefaultmidpunct}
{\mcitedefaultendpunct}{\mcitedefaultseppunct}\relax
\EndOfBibitem
\bibitem[{Dulieu, F.} \latin{et~al.}(2017){Dulieu, F.}, {Minissale, M.}, and
  {Bockel\'ee-Morvan, D.}]{o2.dulieu:2016}
{Dulieu, F.},; {Minissale, M.},; {Bockel\'ee-Morvan, D.}, Production of O$_2$
  through dismutation of H$_2$O$_2$ during water ice desorption: a key to
  understanding comet O$_2$ abundances. \emph{Astron. Astrophys.}
  \textbf{2017}, \emph{597}, A56\relax
\mciteBstWouldAddEndPuncttrue
\mciteSetBstMidEndSepPunct{\mcitedefaultmidpunct}
{\mcitedefaultendpunct}{\mcitedefaultseppunct}\relax
\EndOfBibitem
\bibitem[Pezzella \latin{et~al.}(2018)Pezzella, Unke, and Meuwly]{MM.oxy:2018}
Pezzella,~M.; Unke,~O.~T.; Meuwly,~M. Molecular Oxygen Formation in
  Interstellar Ices Does Not Require Tunneling. \emph{J. Phys. Chem. Lett.}
  \textbf{2018}, \emph{9}, 1822--1826\relax
\mciteBstWouldAddEndPuncttrue
\mciteSetBstMidEndSepPunct{\mcitedefaultmidpunct}
{\mcitedefaultendpunct}{\mcitedefaultseppunct}\relax
\EndOfBibitem
\bibitem[Pezzella and Meuwly(2019)Pezzella, and Meuwly]{MM.oxy:2019}
Pezzella,~M.; Meuwly,~M. O$_2$ formation in cold environments. \emph{Phys.
  Chem. Chem. Phys.} \textbf{2019}, \emph{21}, 6247--6255\relax
\mciteBstWouldAddEndPuncttrue
\mciteSetBstMidEndSepPunct{\mcitedefaultmidpunct}
{\mcitedefaultendpunct}{\mcitedefaultseppunct}\relax
\EndOfBibitem
\bibitem[Christianson and Garrod(2021)Christianson, and
  Garrod]{christianson:2021}
Christianson,~D.~A.; Garrod,~R.~T. Chemical Kinetics Simulations of Ice
  Chemistry on Porous Versus Non-Porous Dust Grains. \emph{Front. astron. space
  sci.} \textbf{2021}, \emph{8}, 643297\relax
\mciteBstWouldAddEndPuncttrue
\mciteSetBstMidEndSepPunct{\mcitedefaultmidpunct}
{\mcitedefaultendpunct}{\mcitedefaultseppunct}\relax
\EndOfBibitem
\bibitem[Hama and Watanabe(2013)Hama, and Watanabe]{hama:2013}
Hama,~T.; Watanabe,~N. Surface Processes on Interstellar Amorphous Solid Water:
  Adsorption, Diffusion, Tunneling Reactions, and Nuclear-Spin Conversion.
  \emph{Chem. Rev.} \textbf{2013}, \emph{113}, 8783--8839\relax
\mciteBstWouldAddEndPuncttrue
\mciteSetBstMidEndSepPunct{\mcitedefaultmidpunct}
{\mcitedefaultendpunct}{\mcitedefaultseppunct}\relax
\EndOfBibitem
\bibitem[Minissale \latin{et~al.}(2013)Minissale, Congiu, Manic\`o, Pirronello,
  and Dulieu]{minissale:2013}
Minissale,~M.; Congiu,~E.; Manic\`o,~G.; Pirronello,~V.; Dulieu,~F. CO$_2$
  formation on interstellar dust grains: a detailed study of the barrier of the
  CO channel. \emph{Astron. Astrophys.} \textbf{2013}, \emph{559}, A49\relax
\mciteBstWouldAddEndPuncttrue
\mciteSetBstMidEndSepPunct{\mcitedefaultmidpunct}
{\mcitedefaultendpunct}{\mcitedefaultseppunct}\relax
\EndOfBibitem
\bibitem[Minissale \latin{et~al.}(2016)Minissale, Moudens, Baouche, Chaabouni,
  and Dulieu]{minissale:2016}
Minissale,~M.; Moudens,~A.; Baouche,~S.; Chaabouni,~H.; Dulieu,~F.
  Hydrogenation of CO-bearing species on grains: unexpected chemical desorption
  of CO. \emph{Mon. Not. Roy. Astron. Soc.} \textbf{2016}, \emph{458},
  2953--2961\relax
\mciteBstWouldAddEndPuncttrue
\mciteSetBstMidEndSepPunct{\mcitedefaultmidpunct}
{\mcitedefaultendpunct}{\mcitedefaultseppunct}\relax
\EndOfBibitem
\bibitem[Qasim \latin{et~al.}(2020)Qasim, Fedoseev, Chuang, He, Ioppolo, van
  Dishoeck, and Linnartz]{qasim:2020}
Qasim,~D.; Fedoseev,~G.; Chuang,~K.-J.; He,~J.; Ioppolo,~S.; van Dishoeck,~E.;
  Linnartz,~H. An experimental study of the surface formation of methane in
  interstellar molecular clouds. \emph{Nat. Astron.} \textbf{2020}, \emph{4},
  781--785\relax
\mciteBstWouldAddEndPuncttrue
\mciteSetBstMidEndSepPunct{\mcitedefaultmidpunct}
{\mcitedefaultendpunct}{\mcitedefaultseppunct}\relax
\EndOfBibitem
\bibitem[Molpeceres \latin{et~al.}(2021)Molpeceres, K\"astner, Fedoseev, Qasim,
  Sch\"omig, Linnartz, and Lamberts]{molpeceres:2021}
Molpeceres,~G.; K\"astner,~J.; Fedoseev,~G.; Qasim,~D.; Sch\"omig,~R.;
  Linnartz,~H.; Lamberts,~T. Carbon Atom Reactivity with Amorphous Solid Water:
  H$_2$O-Catalyzed Formation of H$_2$CO. \emph{J. Phys. Chem. Lett.}
  \textbf{2021}, \emph{12}, 10854--10860\relax
\mciteBstWouldAddEndPuncttrue
\mciteSetBstMidEndSepPunct{\mcitedefaultmidpunct}
{\mcitedefaultendpunct}{\mcitedefaultseppunct}\relax
\EndOfBibitem
\bibitem[Minissale \latin{et~al.}(2014)Minissale, Fedoseev, Congiu, Ioppolo,
  Dulieu, and Linnartz]{no1.minissale:2014}
Minissale,~M.; Fedoseev,~G.; Congiu,~E.; Ioppolo,~S.; Dulieu,~F.; Linnartz,~H.
  Solid state chemistry of nitrogen oxides – Part I: surface consumption of
  NO. \emph{Phys. Chem. Chem. Phys.} \textbf{2014}, \emph{16}, 8257--8269\relax
\mciteBstWouldAddEndPuncttrue
\mciteSetBstMidEndSepPunct{\mcitedefaultmidpunct}
{\mcitedefaultendpunct}{\mcitedefaultseppunct}\relax
\EndOfBibitem
\bibitem[{Minissale, M.} \latin{et~al.}(2019){Minissale, M.}, {Nguyen, T.}, and
  {Dulieu, F.}]{minissale:2018}
{Minissale, M.},; {Nguyen, T.},; {Dulieu, F.}, Experimental study of the
  penetration of oxygen and deuterium atoms into porous water ice.
  \emph{Astron. Astrophys.} \textbf{2019}, \emph{622}, A148\relax
\mciteBstWouldAddEndPuncttrue
\mciteSetBstMidEndSepPunct{\mcitedefaultmidpunct}
{\mcitedefaultendpunct}{\mcitedefaultseppunct}\relax
\EndOfBibitem
\bibitem[Minissale \latin{et~al.}(2016)Minissale, Congiu, and
  Dulieu]{dulieu:2016}
Minissale,~M.; Congiu,~E.; Dulieu,~F. Direct measurement of desorption and
  diffusion energies of O and N atoms physisorbed on amorphous surfaces.
  \emph{Astron. Astrophys.} \textbf{2016}, \emph{585}, A146\relax
\mciteBstWouldAddEndPuncttrue
\mciteSetBstMidEndSepPunct{\mcitedefaultmidpunct}
{\mcitedefaultendpunct}{\mcitedefaultseppunct}\relax
\EndOfBibitem
\bibitem[Tsuge \latin{et~al.}(2020)Tsuge, Hidaka, Kouchi, and
  Watanabe]{tsuge:2020}
Tsuge,~M.; Hidaka,~H.; Kouchi,~A.; Watanabe,~N. Diffusive Hydrogenation
  Reactions of CO Embedded in Amorphous Solid Water at Elevated Temperatures~
  70 K. \emph{Astrophys. J.} \textbf{2020}, \emph{900}, 187\relax
\mciteBstWouldAddEndPuncttrue
\mciteSetBstMidEndSepPunct{\mcitedefaultmidpunct}
{\mcitedefaultendpunct}{\mcitedefaultseppunct}\relax
\EndOfBibitem
\bibitem[Lee and Meuwly(2014)Lee, and Meuwly]{MM.oxy:2014}
Lee,~M.~W.; Meuwly,~M. Diffusion of atomic oxygen relevant to water formation
  in amorphous interstellar ices. \emph{Faraday Discuss.} \textbf{2014},
  \emph{168}, 205--222\relax
\mciteBstWouldAddEndPuncttrue
\mciteSetBstMidEndSepPunct{\mcitedefaultmidpunct}
{\mcitedefaultendpunct}{\mcitedefaultseppunct}\relax
\EndOfBibitem
\bibitem[Roser \latin{et~al.}(2001)Roser, Vidali, Manic{\`o}, and
  Pirronello]{roser:2001}
Roser,~J.~E.; Vidali,~G.; Manic{\`o},~G.; Pirronello,~V. Formation of carbon
  dioxide by surface reactions on ices in the interstellar medium.
  \emph{Astrophys. J. Lett.} \textbf{2001}, \emph{555}, L61\relax
\mciteBstWouldAddEndPuncttrue
\mciteSetBstMidEndSepPunct{\mcitedefaultmidpunct}
{\mcitedefaultendpunct}{\mcitedefaultseppunct}\relax
\EndOfBibitem
\bibitem[Liszt and Turner(1978)Liszt, and Turner]{liszt_NO1978}
Liszt,~H.; Turner,~B. Microwave detection of interstellar NO. \emph{Astrophys.
  J.} \textbf{1978}, \emph{224}, L73--L76\relax
\mciteBstWouldAddEndPuncttrue
\mciteSetBstMidEndSepPunct{\mcitedefaultmidpunct}
{\mcitedefaultendpunct}{\mcitedefaultseppunct}\relax
\EndOfBibitem
\bibitem[Ligterink \latin{et~al.}(2018)Ligterink, Calcutt, Coutens, Kristensen,
  Bourke, Drozdovskaya, M{\"u}ller, Wampfler, van Der~Wiel, Van~Dishoeck,
  \latin{et~al.} others]{ligterink2018alma}
Ligterink,~N. F.~W.; Calcutt,~H.; Coutens,~A.; Kristensen,~L.; Bourke,~T.;
  Drozdovskaya,~M.~N.; M{\"u}ller,~H.; Wampfler,~S.; van Der~Wiel,~M.;
  Van~Dishoeck,~E., \latin{et~al.}  The ALMA-PILS survey: Stringent limits on
  small amines and nitrogen-oxides towards IRAS 16293--2422B. \emph{Astron.
  Astrophys.} \textbf{2018}, \emph{619}, A28\relax
\mciteBstWouldAddEndPuncttrue
\mciteSetBstMidEndSepPunct{\mcitedefaultmidpunct}
{\mcitedefaultendpunct}{\mcitedefaultseppunct}\relax
\EndOfBibitem
\bibitem[Codella \latin{et~al.}(2018)Codella, Viti, Lefloch, Holdship,
  Bachiller, Bianchi, Ceccarelli, Favre, Jim{\'e}nez-Serra, Podio,
  \latin{et~al.} others]{codella2018nitrogen}
Codella,~C.; Viti,~S.; Lefloch,~B.; Holdship,~J.; Bachiller,~R.; Bianchi,~E.;
  Ceccarelli,~C.; Favre,~C.; Jim{\'e}nez-Serra,~I.; Podio,~L., \latin{et~al.}
  Nitrogen oxide in protostellar envelopes and shocks: the ASAI survey.
  \emph{Mon. Notices Royal Astron. Soc.} \textbf{2018}, \emph{474},
  5694--5703\relax
\mciteBstWouldAddEndPuncttrue
\mciteSetBstMidEndSepPunct{\mcitedefaultmidpunct}
{\mcitedefaultendpunct}{\mcitedefaultseppunct}\relax
\EndOfBibitem
\bibitem[Ziurys \latin{et~al.}(1991)Ziurys, McGonagle, Minh, and
  Irvine]{ziurys1991nitric}
Ziurys,~L.; McGonagle,~D.; Minh,~Y.; Irvine,~W. Nitric oxide in star-forming
  regions-Further evidence for interstellar NO bonds. \emph{Astrophys. J.}
  \textbf{1991}, \emph{373}, 535--542\relax
\mciteBstWouldAddEndPuncttrue
\mciteSetBstMidEndSepPunct{\mcitedefaultmidpunct}
{\mcitedefaultendpunct}{\mcitedefaultseppunct}\relax
\EndOfBibitem
\bibitem[McGonagle \latin{et~al.}(1990)McGonagle, Ziurys, Irvine, and
  Minh]{mcgonagle1990detection}
McGonagle,~D.; Ziurys,~L.; Irvine,~W.~M.; Minh,~Y. Detection of nitric oxide in
  the dark cloud L134N. \emph{Astrophys. J.} \textbf{1990}, \emph{359},
  121--124\relax
\mciteBstWouldAddEndPuncttrue
\mciteSetBstMidEndSepPunct{\mcitedefaultmidpunct}
{\mcitedefaultendpunct}{\mcitedefaultseppunct}\relax
\EndOfBibitem
\bibitem[Ziurys \latin{et~al.}(1994)Ziurys, Apponi, Hollis, and
  Snyder]{ziurys1994detection}
Ziurys,~L.; Apponi,~A.; Hollis,~J.; Snyder,~L. Detection of interstellar
  N$_2$O: A new molecule containing an NO bond. \emph{Astrophys. J.}
  \textbf{1994}, \emph{436}, L181--L184\relax
\mciteBstWouldAddEndPuncttrue
\mciteSetBstMidEndSepPunct{\mcitedefaultmidpunct}
{\mcitedefaultendpunct}{\mcitedefaultseppunct}\relax
\EndOfBibitem
\bibitem[Snyder \latin{et~al.}(1993)Snyder, Kuan, Ziurys, and
  Hollis]{snyder1993new}
Snyder,~L.~E.; Kuan,~Y.-J.; Ziurys,~L.; Hollis,~J. New 3 millimeter
  observations of interstellar HNO-Reinstating a discredited identification.
  \emph{Astrophys. J.} \textbf{1993}, \emph{403}, L17--L20\relax
\mciteBstWouldAddEndPuncttrue
\mciteSetBstMidEndSepPunct{\mcitedefaultmidpunct}
{\mcitedefaultendpunct}{\mcitedefaultseppunct}\relax
\EndOfBibitem
\bibitem[de~Barros \latin{et~al.}(2016)de~Barros, Da~Silveira, Fulvio, Boduch,
  and Rothard]{de2016formation}
de~Barros,~A.; Da~Silveira,~E.; Fulvio,~D.; Boduch,~P.; Rothard,~H. Formation
  of nitrogen-and oxygen-bearing molecules from radiolysis of nitrous oxide
  ices--implications for Solar system and interstellar ices. \emph{Mon. Notices
  Royal Astron. Soc.} \textbf{2016}, \emph{465}, 3281--3290\relax
\mciteBstWouldAddEndPuncttrue
\mciteSetBstMidEndSepPunct{\mcitedefaultmidpunct}
{\mcitedefaultendpunct}{\mcitedefaultseppunct}\relax
\EndOfBibitem
\bibitem[Congiu \latin{et~al.}(2012)Congiu, Fedoseev, Ioppolo, Dulieu,
  Chaabouni, Baouche, Lemaire, Laffon, Parent, Lamberts, \latin{et~al.}
  others]{congiu2012no}
Congiu,~E.; Fedoseev,~G.; Ioppolo,~S.; Dulieu,~F.; Chaabouni,~H.; Baouche,~S.;
  Lemaire,~J.~L.; Laffon,~C.; Parent,~P.; Lamberts,~T., \latin{et~al.}  No ice
  hydrogenation: a solid pathway to NH$_2$OH formation in space.
  \emph{Astrophys. J. Lett.} \textbf{2012}, \emph{750}, L12\relax
\mciteBstWouldAddEndPuncttrue
\mciteSetBstMidEndSepPunct{\mcitedefaultmidpunct}
{\mcitedefaultendpunct}{\mcitedefaultseppunct}\relax
\EndOfBibitem
\bibitem[Stief \latin{et~al.}(1975)Stief, Payne, and Klemm]{klemm:1975}
Stief,~L.~J.; Payne,~W.~A.; Klemm,~R.~B. A flash photolysis--resonance
  fluorescence study of the formation of O($^1$D) in the photolysis of water
  and the reaction of O($^1$D) with H$_2$, Ar, and He-. \emph{J. Chem. Phys.}
  \textbf{1975}, \emph{62}, 4000--4008\relax
\mciteBstWouldAddEndPuncttrue
\mciteSetBstMidEndSepPunct{\mcitedefaultmidpunct}
{\mcitedefaultendpunct}{\mcitedefaultseppunct}\relax
\EndOfBibitem
\bibitem[Garstang(1951)]{garstang:1951}
Garstang,~R. Energy levels and transition probabilities in p$^2$ and p$^4$
  configurations. \emph{Mon. Not. R. Astron. Soc.} \textbf{1951}, \emph{111},
  115--124\relax
\mciteBstWouldAddEndPuncttrue
\mciteSetBstMidEndSepPunct{\mcitedefaultmidpunct}
{\mcitedefaultendpunct}{\mcitedefaultseppunct}\relax
\EndOfBibitem
\bibitem[Schmidt \latin{et~al.}(2019)Schmidt, Swiderek, and
  Bredeho\"oft]{schmidt:2019}
Schmidt,~F.; Swiderek,~P.; Bredeho\"oft,~J.~H. Formation of Formic Acid,
  Formaldehyde, and Carbon Dioxide by Electron-Induced Chemistry in Ices of
  Water and Carbon Monoxide. \emph{ACS Earth Space Chem.} \textbf{2019},
  \emph{3}, 1974--1986\relax
\mciteBstWouldAddEndPuncttrue
\mciteSetBstMidEndSepPunct{\mcitedefaultmidpunct}
{\mcitedefaultendpunct}{\mcitedefaultseppunct}\relax
\EndOfBibitem
\bibitem[Chou \latin{et~al.}(2018)Chou, Lo, Peng, Lu, Cheng, and
  Ogilvie]{ogilvie:2018}
Chou,~S.-L.; Lo,~J.-I.; Peng,~Y.-C.; Lu,~H.-C.; Cheng,~B.-M.; Ogilvie,~J.~F.
  Photolysis of O$_2$ dispersed in solid neon with far-ultraviolet radiation.
  \emph{Phys. Chem. Chem. Phys.} \textbf{2018}, \emph{20}, 7730--7738\relax
\mciteBstWouldAddEndPuncttrue
\mciteSetBstMidEndSepPunct{\mcitedefaultmidpunct}
{\mcitedefaultendpunct}{\mcitedefaultseppunct}\relax
\EndOfBibitem
\bibitem[T.~Nagy and Meuwly(2014)T.~Nagy, and Meuwly]{msarmd}
T.~Nagy,~J. Y.~R.; Meuwly,~M. Multi-Surface Adiabatic Reactive Molecular
  Dynamics. \emph{J. Chem. Theo. Comp.} \textbf{2014}, \emph{10},
  1366--1375\relax
\mciteBstWouldAddEndPuncttrue
\mciteSetBstMidEndSepPunct{\mcitedefaultmidpunct}
{\mcitedefaultendpunct}{\mcitedefaultseppunct}\relax
\EndOfBibitem
\bibitem[Werner \latin{et~al.}(2020)Werner, Knowles, Manby, Black, Doll,
  Hesselmann, Kats, Koehn, Korona, Kreplin, Ma, Miller, Mitrushchenkov,
  Peterson, Polyak, Rauhut, and Sibaev]{molpro:2020}
Werner,~H.-J. \latin{et~al.}  The Molpro quantum chemistry package. \emph{J.
  Chem. Phys.} \textbf{2020}, \emph{152}, 144107\relax
\mciteBstWouldAddEndPuncttrue
\mciteSetBstMidEndSepPunct{\mcitedefaultmidpunct}
{\mcitedefaultendpunct}{\mcitedefaultseppunct}\relax
\EndOfBibitem
\bibitem[Werner and Knowles(1988)Werner, and Knowles]{werner1988efficient}
Werner,~H.-J.; Knowles,~P.~J. An efficient internally contracted
  multiconfiguration--reference configuration interaction method. \emph{J.
  Chem. Phys.} \textbf{1988}, \emph{89}, 5803--5814\relax
\mciteBstWouldAddEndPuncttrue
\mciteSetBstMidEndSepPunct{\mcitedefaultmidpunct}
{\mcitedefaultendpunct}{\mcitedefaultseppunct}\relax
\EndOfBibitem
\bibitem[Dunning~Jr(1989)]{dunning1989gaussian}
Dunning~Jr,~T.~H. Gaussian basis sets for use in correlated molecular
  calculations. I. The atoms boron through neon and hydrogen. \emph{J. Chem.
  Phys.} \textbf{1989}, \emph{90}, 1007--1023\relax
\mciteBstWouldAddEndPuncttrue
\mciteSetBstMidEndSepPunct{\mcitedefaultmidpunct}
{\mcitedefaultendpunct}{\mcitedefaultseppunct}\relax
\EndOfBibitem
\bibitem[Best \latin{et~al.}(2012)Best, Zhu, Shim, Lopes, Mittal, Feig, and
  MacKerell~Jr]{best2012optimization}
Best,~R.~B.; Zhu,~X.; Shim,~J.; Lopes,~P.~E.; Mittal,~J.; Feig,~M.;
  MacKerell~Jr,~A.~D. Optimization of the additive CHARMM all-atom protein
  force field targeting improved sampling of the backbone $\phi$, $\psi$ and
  side-chain $\chi_1$ and $\chi_2$ dihedral angles. \emph{J. Chem. Theo. Comp.}
  \textbf{2012}, \emph{8}, 3257--3273\relax
\mciteBstWouldAddEndPuncttrue
\mciteSetBstMidEndSepPunct{\mcitedefaultmidpunct}
{\mcitedefaultendpunct}{\mcitedefaultseppunct}\relax
\EndOfBibitem
\bibitem[Pezzella \latin{et~al.}(2020)Pezzella, Koner, and Meuwly]{MM.o2:2020}
Pezzella,~M.; Koner,~D.; Meuwly,~M. Formation and Stabilization of Ground and
  Excited-State Singlet O$_2$ upon Recombination of $^3$P Oxygen on Amorphous
  Solid Water. \emph{J. Phys. Chem. Lett.} \textbf{2020}, \emph{11},
  2171--2176\relax
\mciteBstWouldAddEndPuncttrue
\mciteSetBstMidEndSepPunct{\mcitedefaultmidpunct}
{\mcitedefaultendpunct}{\mcitedefaultseppunct}\relax
\EndOfBibitem
\bibitem[Upadhyay \latin{et~al.}(2021)Upadhyay, Pezzella, and
  Meuwly]{MM.genesis:2021}
Upadhyay,~M.; Pezzella,~M.; Meuwly,~M. Genesis of Polyatomic Molecules in Dark
  Clouds: CO$_2$ Formation on Cold Amorphous Solid Water. \emph{J. Phys. Chem.
  Lett.} \textbf{2021}, \emph{12}, 6781--6787\relax
\mciteBstWouldAddEndPuncttrue
\mciteSetBstMidEndSepPunct{\mcitedefaultmidpunct}
{\mcitedefaultendpunct}{\mcitedefaultseppunct}\relax
\EndOfBibitem
\bibitem[San Vicente~Veliz \latin{et~al.}(2020)San Vicente~Veliz, Koner,
  Schwilk, Bemish, and Meuwly]{MM.NO2.2020}
San Vicente~Veliz,~J.~C.; Koner,~D.; Schwilk,~M.; Bemish,~R.~J.; Meuwly,~M. The
  N($^4$S) + O$_2$(X$^{3} \Sigma$) $\longleftrightarrow$ O($^3$P) +
  NO(X$^{2}\Pi$) Reaction: Thermal and Vibrational Relaxation Rates for the
  $^2$A', $^4$A' and $^2$A'' States. \emph{Phys. Chem. Chem. Phys.}
  \textbf{2020}, \emph{22}, 3927--3939\relax
\mciteBstWouldAddEndPuncttrue
\mciteSetBstMidEndSepPunct{\mcitedefaultmidpunct}
{\mcitedefaultendpunct}{\mcitedefaultseppunct}\relax
\EndOfBibitem
\bibitem[Brooks \latin{et~al.}(2009)Brooks, III, Jr, Nilsson, Petrella, Roux,
  Won, Archontis, Bartels, Boresch, Caflisch, Caves, Cui, Dinner, Feig,
  Fischer, Gao, Hodoscek, Im, Kuczera, Lazaridis, Ma, Ovchinnikov, Paci,
  Pastor, Post, Pu, Schaefer, Tidor, Venable, Woodcock, Wu, Yang, York, and
  Karplus]{charmm.prog}
Brooks,~B. \latin{et~al.}  CHARMM: the biomolecular simulation program.
  \emph{J. Comp. Chem.} \textbf{2009}, \emph{30}, 1545--614\relax
\mciteBstWouldAddEndPuncttrue
\mciteSetBstMidEndSepPunct{\mcitedefaultmidpunct}
{\mcitedefaultendpunct}{\mcitedefaultseppunct}\relax
\EndOfBibitem
\bibitem[Unke and Meuwly(2017)Unke, and Meuwly]{MM.rkhs:2017}
Unke,~O.~T.; Meuwly,~M. Toolkit for the Construction of Reproducing
  Kernel-Based Representations of Data: Application to Multidimensional
  Potential Energy Surfaces. \emph{J. Chem. Inf. Model.} \textbf{2017},
  \emph{57}, 1923--1931\relax
\mciteBstWouldAddEndPuncttrue
\mciteSetBstMidEndSepPunct{\mcitedefaultmidpunct}
{\mcitedefaultendpunct}{\mcitedefaultseppunct}\relax
\EndOfBibitem
\bibitem[Gaigeot \latin{et~al.}(2007)Gaigeot, Martinez, and
  Vuilleumier]{Gaigeot2007vdos}
Gaigeot,~M.-P.; Martinez,~M.; Vuilleumier,~R. Infrared spectroscopy in the gas
  and liquid phase from first principle molecular dynamics simulations:
  application to small peptides. \emph{Mol. Phys.} \textbf{2007}, \emph{105},
  2857--2878\relax
\mciteBstWouldAddEndPuncttrue
\mciteSetBstMidEndSepPunct{\mcitedefaultmidpunct}
{\mcitedefaultendpunct}{\mcitedefaultseppunct}\relax
\EndOfBibitem
\bibitem[Ferrero \latin{et~al.}(2023)Ferrero, Pantaleone, Ceccarelli, Ugliengo,
  Sodupe, and Rimola]{ferrero2023NH3vdos}
Ferrero,~S.; Pantaleone,~S.; Ceccarelli,~C.; Ugliengo,~P.; Sodupe,~M.;
  Rimola,~A. Where does the energy go during the interstellar NH$_3$ formation
  on water ice? A computational study. \emph{Astrophys. J.} \textbf{2023},
  \emph{944}, 142\relax
\mciteBstWouldAddEndPuncttrue
\mciteSetBstMidEndSepPunct{\mcitedefaultmidpunct}
{\mcitedefaultendpunct}{\mcitedefaultseppunct}\relax
\EndOfBibitem
\bibitem[Futrelle and McGinty(1971)Futrelle, and McGinty]{futrelle1971vdos}
Futrelle,~R.; McGinty,~D. Calculation of spectra and correlation functions from
  molecular dynamics data using the fast Fourier transform. \emph{Chemical
  Physics Letters} \textbf{1971}, \emph{12}, 285--287\relax
\mciteBstWouldAddEndPuncttrue
\mciteSetBstMidEndSepPunct{\mcitedefaultmidpunct}
{\mcitedefaultendpunct}{\mcitedefaultseppunct}\relax
\EndOfBibitem
\bibitem[Burnham \latin{et~al.}(1997)Burnham, Li, and Leslie]{Burnham97p6192}
Burnham,~C.~J.; Li,~J.~C.; Leslie,~M. Molecular Dynamics Calculations for Ice
  {I}h. \emph{J. Phys. Chem. B} \textbf{1997}, \emph{101}, 6192--6195\relax
\mciteBstWouldAddEndPuncttrue
\mciteSetBstMidEndSepPunct{\mcitedefaultmidpunct}
{\mcitedefaultendpunct}{\mcitedefaultseppunct}\relax
\EndOfBibitem
\bibitem[Plattner and Meuwly(2008)Plattner, and Meuwly]{MM.ice:2008}
Plattner,~N.; Meuwly,~M. Atomistic Simulations of CO Vibrations in Ices
  Relevant to Astrochemistry. \emph{ChemPhysChem} \textbf{2008}, \emph{9},
  1271--1277\relax
\mciteBstWouldAddEndPuncttrue
\mciteSetBstMidEndSepPunct{\mcitedefaultmidpunct}
{\mcitedefaultendpunct}{\mcitedefaultseppunct}\relax
\EndOfBibitem
\bibitem[Kumagai \latin{et~al.}(1994)Kumagai, Kawamura, and Yokokawa]{kky_orig}
Kumagai,~N.; Kawamura,~K.; Yokokawa,~T. An Interatomic Potential Model for
  H$_2$O: Applications to Water and Ice Polymorphs. \emph{Mol. Sim.}
  \textbf{1994}, \emph{12}, 177--186\relax
\mciteBstWouldAddEndPuncttrue
\mciteSetBstMidEndSepPunct{\mcitedefaultmidpunct}
{\mcitedefaultendpunct}{\mcitedefaultseppunct}\relax
\EndOfBibitem
\bibitem[Lee and Meuwly(2011)Lee, and Meuwly]{MM.cn:2011}
Lee,~M.~W.; Meuwly,~M. On the role of nonbonded interactions in vibrational
  energy relaxation of cyanide in water. \emph{J. Phys. Chem. A} \textbf{2011},
  \emph{115}, 5053--5061\relax
\mciteBstWouldAddEndPuncttrue
\mciteSetBstMidEndSepPunct{\mcitedefaultmidpunct}
{\mcitedefaultendpunct}{\mcitedefaultseppunct}\relax
\EndOfBibitem
\bibitem[Yu \latin{et~al.}(2020)Yu, Chiang, Okuno, Seki, Ohto, Yu, Korepanov,
  Hamaguchi, Bonn, Hunger, \latin{et~al.} others]{yu:2020}
Yu,~C.-C.; Chiang,~K.-Y.; Okuno,~M.; Seki,~T.; Ohto,~T.; Yu,~X.; Korepanov,~V.;
  Hamaguchi,~H.-o.; Bonn,~M.; Hunger,~J., \latin{et~al.}  Vibrational couplings
  and energy transfer pathways of water’s bending mode. \emph{Nat. Commun.}
  \textbf{2020}, \emph{11}, 5977\relax
\mciteBstWouldAddEndPuncttrue
\mciteSetBstMidEndSepPunct{\mcitedefaultmidpunct}
{\mcitedefaultendpunct}{\mcitedefaultseppunct}\relax
\EndOfBibitem
\bibitem[Devlin(1990)]{devlin:1990}
Devlin,~J.~P. Vibrational modes of amorphous ice: bending mode frequencies for
  isotopically decoupled H$_2$O and HOD at 90 K. \emph{J. Mol. Struct.}
  \textbf{1990}, \emph{224}, 33--43\relax
\mciteBstWouldAddEndPuncttrue
\mciteSetBstMidEndSepPunct{\mcitedefaultmidpunct}
{\mcitedefaultendpunct}{\mcitedefaultseppunct}\relax
\EndOfBibitem
\bibitem[Shimanouchi \latin{et~al.}(1978)Shimanouchi, Matsuura, Ogawa, and
  Harada]{shimanouchi:1978}
Shimanouchi,~T.; Matsuura,~H.; Ogawa,~Y.; Harada,~I. Tables of molecular
  vibrational frequencies. \emph{J. Phys. Chem. Ref. Data} \textbf{1978},
  \emph{7}, 1323--1444\relax
\mciteBstWouldAddEndPuncttrue
\mciteSetBstMidEndSepPunct{\mcitedefaultmidpunct}
{\mcitedefaultendpunct}{\mcitedefaultseppunct}\relax
\EndOfBibitem
\bibitem[Arakawa and Nielsen(1958)Arakawa, and Nielsen]{arakawa:1958_NO2-IR}
Arakawa,~E.~T.; Nielsen,~A.~H. Infrared spectra and molecular constants of
  N$^{14}$O$_2$ and N$^{15}$O$_2$. \emph{J. Mol. Spectrosc.} \textbf{1958},
  \emph{2}, 413--427\relax
\mciteBstWouldAddEndPuncttrue
\mciteSetBstMidEndSepPunct{\mcitedefaultmidpunct}
{\mcitedefaultendpunct}{\mcitedefaultseppunct}\relax
\EndOfBibitem
\bibitem[Karssemeijer \latin{et~al.}(2013)Karssemeijer, Ioppolo, van Hemert,
  van~der Avoird, Allodi, Blake, and Cuppen]{cuppen:2013}
Karssemeijer,~L.; Ioppolo,~S.; van Hemert,~M.; van~der Avoird,~A.; Allodi,~M.;
  Blake,~G.; Cuppen,~H. Dynamics of CO in amorphous water-ice environments.
  \emph{Astrophys. J.} \textbf{2013}, \emph{781}, 1--15\relax
\mciteBstWouldAddEndPuncttrue
\mciteSetBstMidEndSepPunct{\mcitedefaultmidpunct}
{\mcitedefaultendpunct}{\mcitedefaultseppunct}\relax
\EndOfBibitem
\bibitem[Noble \latin{et~al.}(2012)Noble, Congiu, Dulieu, and
  Fraser]{noble:2012}
Noble,~J.; Congiu,~E.; Dulieu,~F.; Fraser,~H. Thermal desorption
  characteristics of CO, O$_2$ and CO$_2$ on non-porous water, crystalline
  water and silicate surfaces at submonolayer and multilayer coverages.
  \emph{Mon. Not. R. Astron. Soc.} \textbf{2012}, \emph{421}, 768--779\relax
\mciteBstWouldAddEndPuncttrue
\mciteSetBstMidEndSepPunct{\mcitedefaultmidpunct}
{\mcitedefaultendpunct}{\mcitedefaultseppunct}\relax
\EndOfBibitem
\bibitem[Minissale \latin{et~al.}(2019)Minissale, Nguyen, and
  Dulieu]{minissale2019experimental}
Minissale,~M.; Nguyen,~T.; Dulieu,~F. Experimental study of the penetration of
  oxygen and deuterium atoms into porous water ice. \emph{Astron. Astrophys.}
  \textbf{2019}, \emph{622}, A148\relax
\mciteBstWouldAddEndPuncttrue
\mciteSetBstMidEndSepPunct{\mcitedefaultmidpunct}
{\mcitedefaultendpunct}{\mcitedefaultseppunct}\relax
\EndOfBibitem
\bibitem[Galvez \latin{et~al.}(2007)Galvez, Ortega, Mat{\'e}, Moreno,
  Mart{\'\i}n-Llorente, Herrero, Escribano, and
  Guti{\'e}rrez]{galvez2007co2desorption}
Galvez,~O.; Ortega,~I.~K.; Mat{\'e},~B.; Moreno,~M.~A.;
  Mart{\'\i}n-Llorente,~B.; Herrero,~V.~J.; Escribano,~R.; Guti{\'e}rrez,~P.~J.
  A study of the interaction of CO$_2$ with water ice. \emph{Astron.
  Astrophys.} \textbf{2007}, \emph{472}, 691--698\relax
\mciteBstWouldAddEndPuncttrue
\mciteSetBstMidEndSepPunct{\mcitedefaultmidpunct}
{\mcitedefaultendpunct}{\mcitedefaultseppunct}\relax
\EndOfBibitem
\bibitem[Wakelam \latin{et~al.}(2017)Wakelam, Loison, Mereau, and
  Ruaud]{wakelam.binding:2017}
Wakelam,~V.; Loison,~J.-C.; Mereau,~R.; Ruaud,~M. Binding energies: New values
  and impact on the efficiency of chemical desorption. \emph{Mol. Astrophys.}
  \textbf{2017}, \emph{6}, 22--35\relax
\mciteBstWouldAddEndPuncttrue
\mciteSetBstMidEndSepPunct{\mcitedefaultmidpunct}
{\mcitedefaultendpunct}{\mcitedefaultseppunct}\relax
\EndOfBibitem
\bibitem[Ioppolo \latin{et~al.}(2014)Ioppolo, Fedoseev, Minissale, Congiu,
  Dulieu, and Linnartz]{linnartz:2014-2}
Ioppolo,~S.; Fedoseev,~G.; Minissale,~M.; Congiu,~E.; Dulieu,~F.; Linnartz,~H.
  Solid state chemistry of nitrogen oxides--Part II: surface consumption of
  NO$_2$. \emph{Phys. Chem. Chem. Phys.} \textbf{2014}, \emph{16},
  8270--8282\relax
\mciteBstWouldAddEndPuncttrue
\mciteSetBstMidEndSepPunct{\mcitedefaultmidpunct}
{\mcitedefaultendpunct}{\mcitedefaultseppunct}\relax
\EndOfBibitem
\bibitem[Acharyya(2022)]{acharyya:2022}
Acharyya,~K. Understanding the impact of diffusion of CO in the astrochemical
  models. \emph{Pub. Astron. Soc. Austr.} \textbf{2022}, \emph{39}, e009\relax
\mciteBstWouldAddEndPuncttrue
\mciteSetBstMidEndSepPunct{\mcitedefaultmidpunct}
{\mcitedefaultendpunct}{\mcitedefaultseppunct}\relax
\EndOfBibitem
\bibitem[Karssemeijer and Cuppen(2014)Karssemeijer, and Cuppen]{cuppen:2014}
Karssemeijer,~L.; Cuppen,~H. Diffusion-desorption ratio of adsorbed CO and
  CO$_2$ on water ice. \emph{Astron. Astrophys.} \textbf{2014}, \emph{569},
  A107\relax
\mciteBstWouldAddEndPuncttrue
\mciteSetBstMidEndSepPunct{\mcitedefaultmidpunct}
{\mcitedefaultendpunct}{\mcitedefaultseppunct}\relax
\EndOfBibitem
\bibitem[He \latin{et~al.}(2018)He, Emtiaz, and Vidali]{he:2018}
He,~J.; Emtiaz,~S.; Vidali,~G. Measurements of Diffusion of Volatiles in
  Amorphous Solid Water: Application to Interstellar Medium Environments.
  \emph{Astrophys. J.} \textbf{2018}, \emph{863}, 156\relax
\mciteBstWouldAddEndPuncttrue
\mciteSetBstMidEndSepPunct{\mcitedefaultmidpunct}
{\mcitedefaultendpunct}{\mcitedefaultseppunct}\relax
\EndOfBibitem
\bibitem[Fredon \latin{et~al.}(2017)Fredon, Lamberts, and
  Cuppen]{cuppen:2017_nonthermaldiffusion}
Fredon,~A.; Lamberts,~T.; Cuppen,~H. Energy dissipation and nonthermal
  diffusion on interstellar ice grains. \emph{Astrophys. J.} \textbf{2017},
  \emph{849}, 125\relax
\mciteBstWouldAddEndPuncttrue
\mciteSetBstMidEndSepPunct{\mcitedefaultmidpunct}
{\mcitedefaultendpunct}{\mcitedefaultseppunct}\relax
\EndOfBibitem
\bibitem[Upadhyay \latin{et~al.}(2023)Upadhyay, Topfer, and
  Meuwly]{MM.criegee:2023}
Upadhyay,~M.; Topfer,~K.; Meuwly,~M. Molecular Simulation for Atmospheric
  Reactions: Non-Equilibrium Dynamics, Roaming, an Glycolaldehyde Formation
  following Photoinduced Decomposition of syn-Acetaldehyde Oxide. \emph{J.
  Phys. Chem. Lett.} \textbf{2023}, \emph{15}, 90--96\relax
\mciteBstWouldAddEndPuncttrue
\mciteSetBstMidEndSepPunct{\mcitedefaultmidpunct}
{\mcitedefaultendpunct}{\mcitedefaultseppunct}\relax
\EndOfBibitem
\end{mcitethebibliography}


\providecommand{\latin}[1]{#1}
\makeatletter
\providecommand{\doi}
  {\begingroup\let\do\@makeother\dospecials
  \catcode`\{=1 \catcode`\}=2 \doi@aux}
\providecommand{\doi@aux}[1]{\endgroup\texttt{#1}}
\makeatother
\providecommand*\mcitethebibliography{\thebibliography}
\csname @ifundefined\endcsname{endmcitethebibliography}
  {\let\endmcitethebibliography\endthebibliography}{}
\begin{mcitethebibliography}{0}
\providecommand*\natexlab[1]{#1}
\providecommand*\mciteSetBstSublistMode[1]{}
\providecommand*\mciteSetBstMaxWidthForm[2]{}
\providecommand*\mciteBstWouldAddEndPuncttrue
  {\def\EndOfBibitem{\unskip.}}
\providecommand*\mciteBstWouldAddEndPunctfalse
  {\let\EndOfBibitem\relax}
\providecommand*\mciteSetBstMidEndSepPunct[3]{}
\providecommand*\mciteSetBstSublistLabelBeginEnd[3]{}
\providecommand*\EndOfBibitem{}
\mciteSetBstSublistMode{f}
\mciteSetBstMaxWidthForm{subitem}{(\alph{mcitesubitemcount})}
\mciteSetBstSublistLabelBeginEnd
  {\mcitemaxwidthsubitemform\space}
  {\relax}
  {\relax}

\end{mcitethebibliography}

\end{document}


\date{\today}

\begin{figure}[h!]
\centering \includegraphics[scale=0.28]{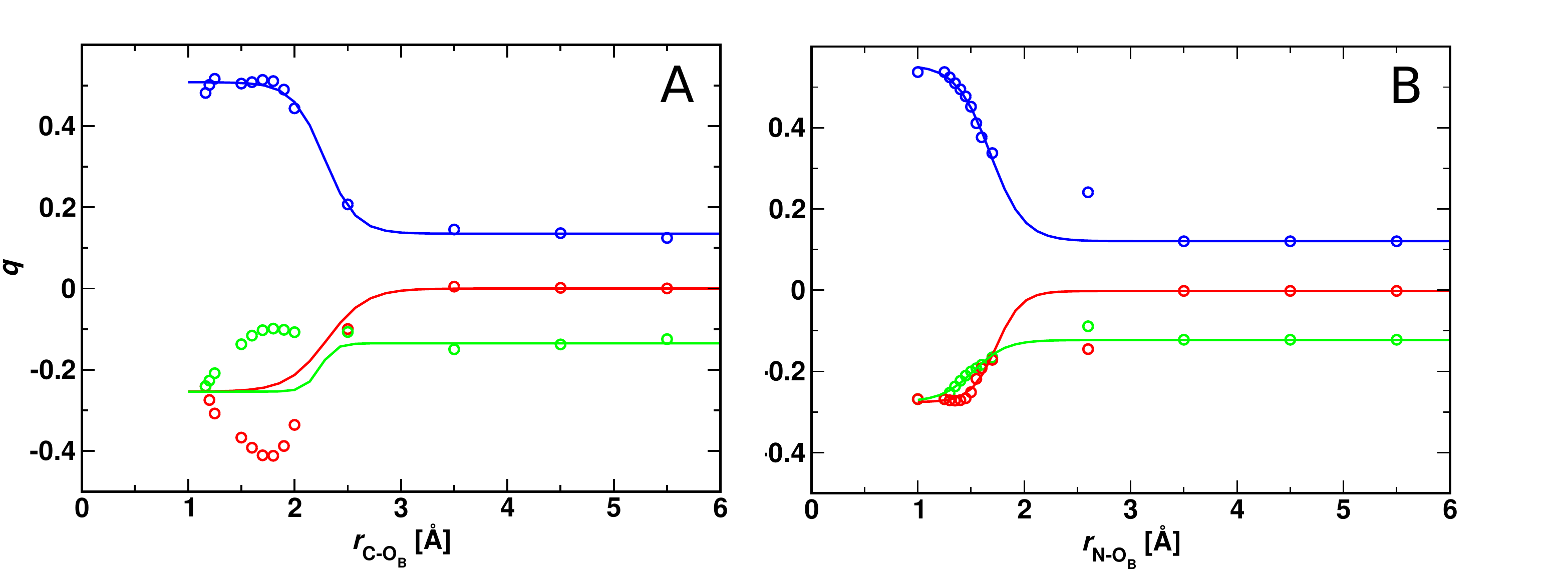}
 \caption{Change in charges of the atoms adsorbed to the 10 nearest
   H$_2$O molecules during CO$_2$/NO$_2$ formation as a function of
   $r$(C/N--O$\rm _B$) \AA\/ distance. Panels A and B correspond to
   CO$_2$ and NO$_2$, respectively. At low temperatures, the reaction
   probability completely depends on the diffusion of species for
   which charges fit well at asymptotic. At close range ($r \sim 2$
   \AA\/ and shorter) bonded interactions dominate. Finally, at
   equilibrium, correct charges are achieved by the fitted
   function. Filled circles and solid line refers to {\it ab initio}
   determined and fitted charges, respectively. Color code: C/N atoms
   (blue) and atomic oxygen O$\rm _B$ (red).}
\label{sifig:charges}
\end{figure}

\begin{figure}[h!]
\centering \includegraphics[scale=0.90]{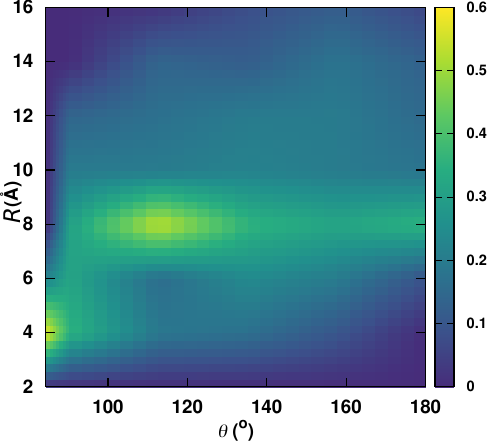}
\caption{KDE density map for the average COO formation probability
  depending on initial $(R,\theta)$ using RKHS PES. }
   \label{sifig:coo}
\end{figure}

\begin{figure}
    \centering
    \includegraphics[scale=0.40]{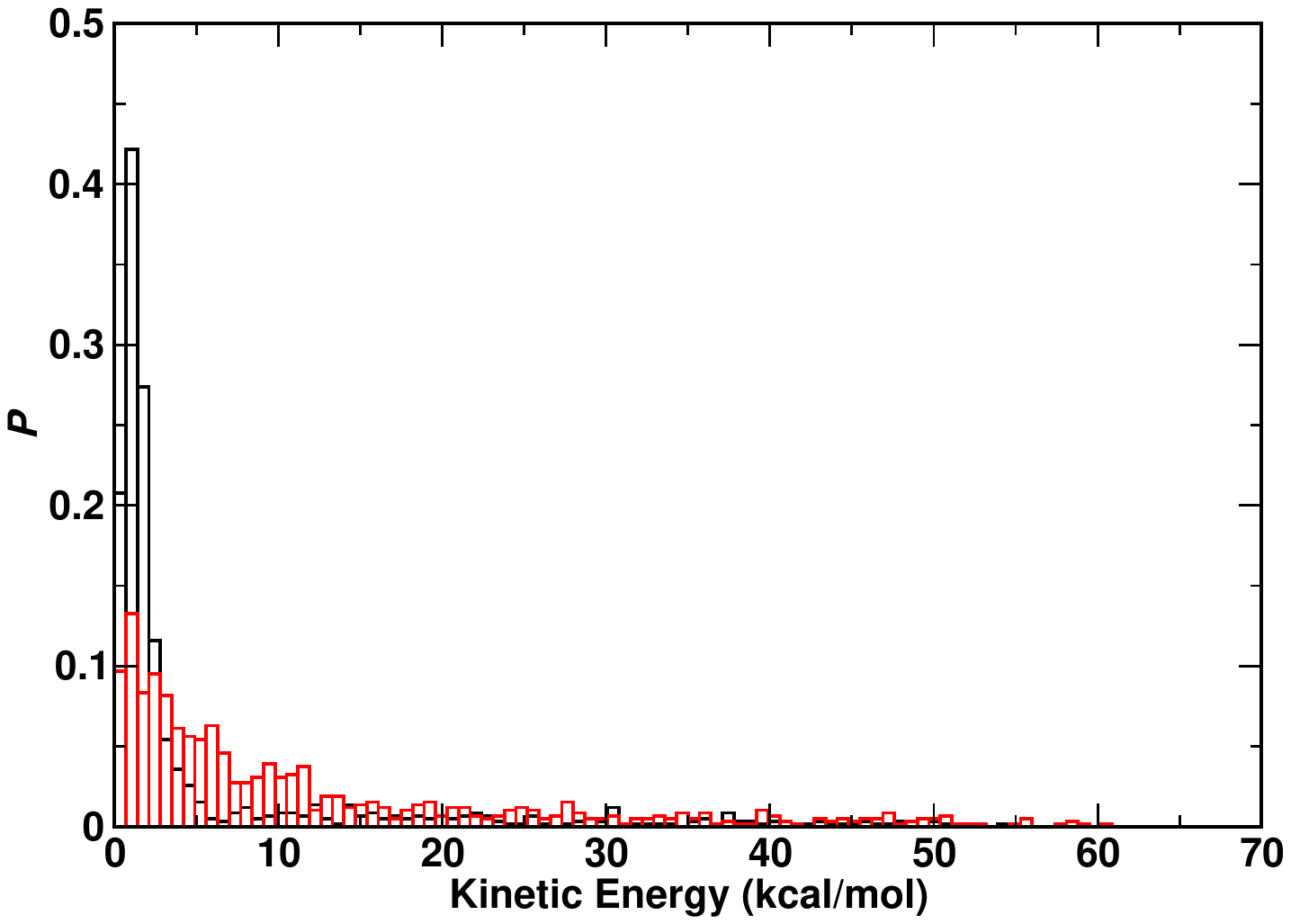}   
    \caption{Normalized kinetic energy distribution for CO$_2$ (red)
      and NO$_2$ (black) from 280 reactive trajectories after 450 ps
      of recombination using RKHS PES. For NO$_2$ $\sim 80$ \% of
      trajectories relaxed down to $\sim 5$ kcal/mol whereas this
      fraction is only $40$ \% for CO$_2$. In other words, upon
      recombination NO$_2$ relaxes more effectively than CO$_2$ on the
      sub-ns time scale.}
    \label{sifig:ke}
\end{figure}

\begin{figure}[h!]
\centering \includegraphics[scale=0.35]{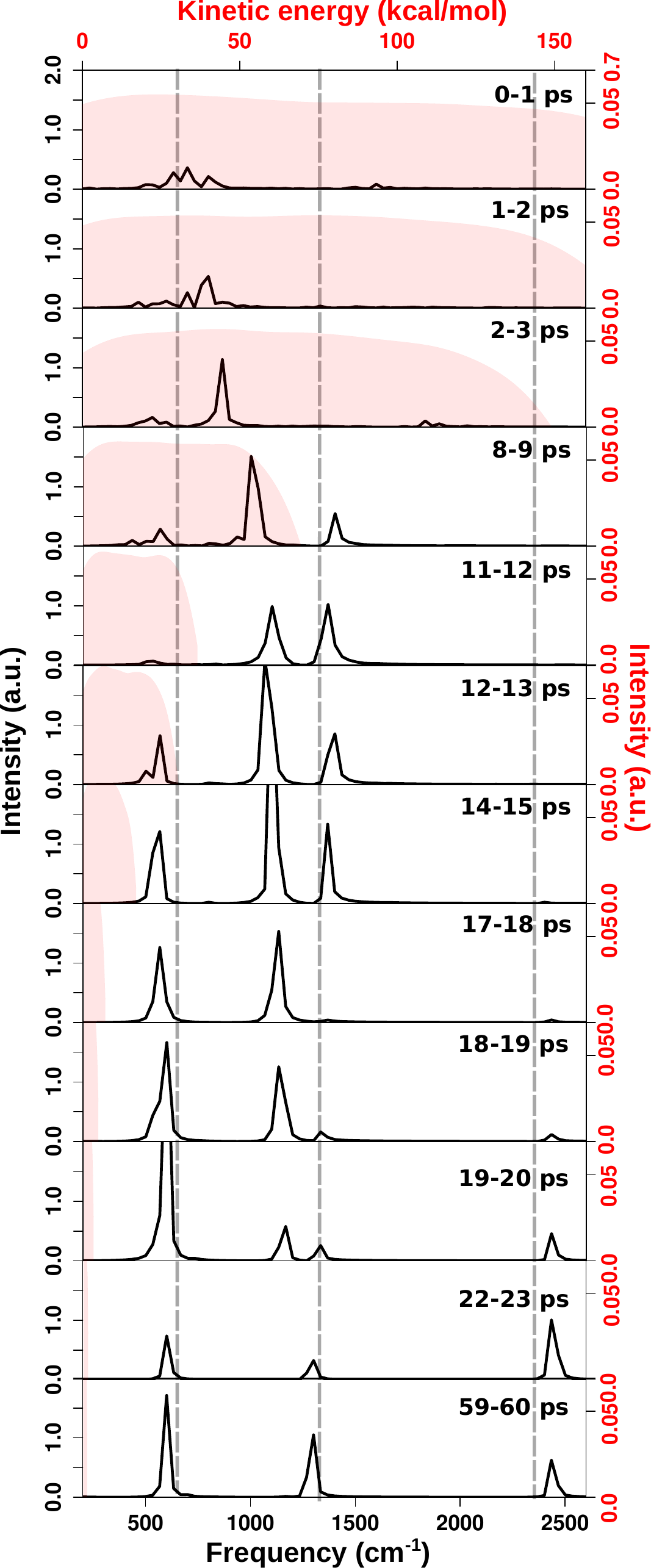}
\caption{The vibrational density of states power spectrum (black) and kinetic energy 
distribution (red) of CO$_2$ after recombination. }
\label{sifig:blowup-co2-vdos}
\end{figure}

\begin{figure}[h!]
\centering
    \includegraphics[scale=0.45]{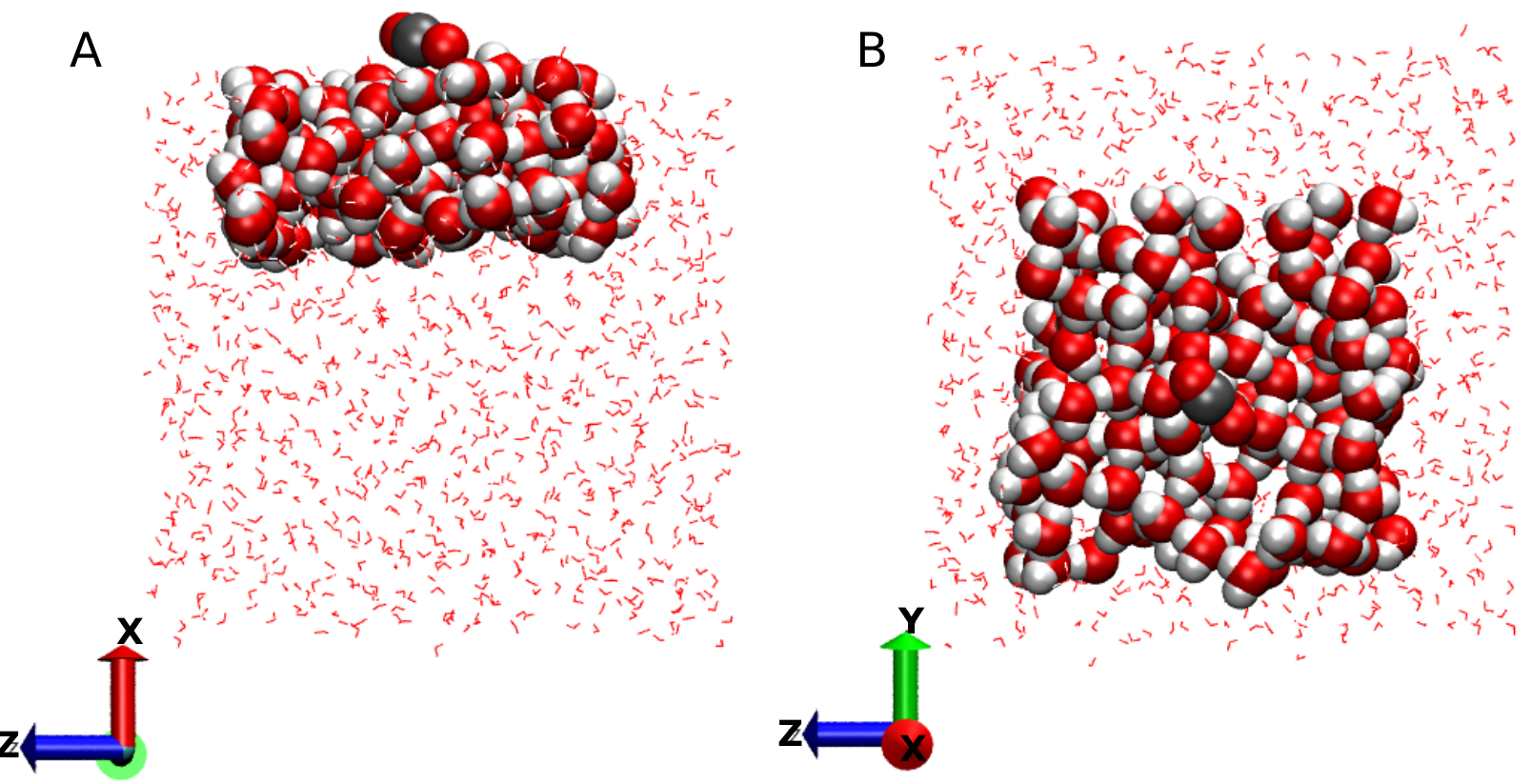}
    \caption{Side (left) and top (right) view of the ASW surface with
      CO$_2$ on the top. The water molecules within 10 \AA\/ in the
      $x-$direction and 5 \AA\/ in the $y-/z-$directions are
      highlighted.}
    \label{sifig:water-xyz}
\end{figure}

\begin{figure}[h!]
\centering \includegraphics[scale=0.90]{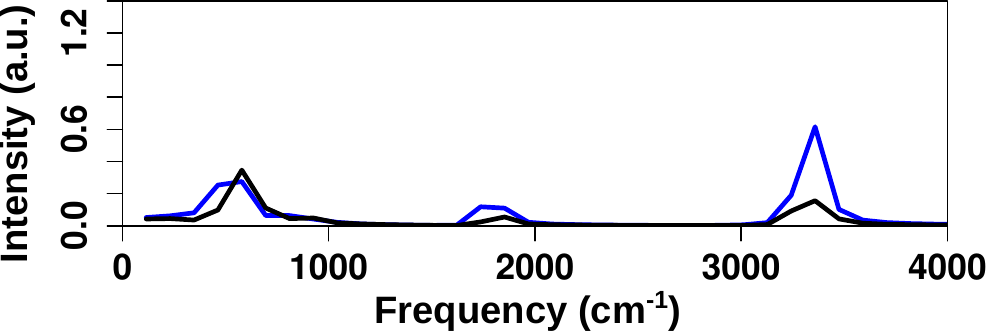}
\caption{The vDOS spectrum for 10 randomly chosen surface water
  molecules (including W$_{\rm A}$ and W$_{\rm B}$) before formation
  of CO$_2$ (black, thermal equilibrium) and 120 ps after CO+O
  recombination (blue). The spectrum at $t= 120$ ps reproduces all
  features of the equilibrium vDOS, i.e. the system has returned to
  equilibrium on the 120 ps time scale. The intensities are
  ``populations of modes within a given frequency interval''.}
\label{sifig:vdos-bef-aft}
\end{figure}

\begin{figure}[h!]
\centering \includegraphics[scale=0.70]{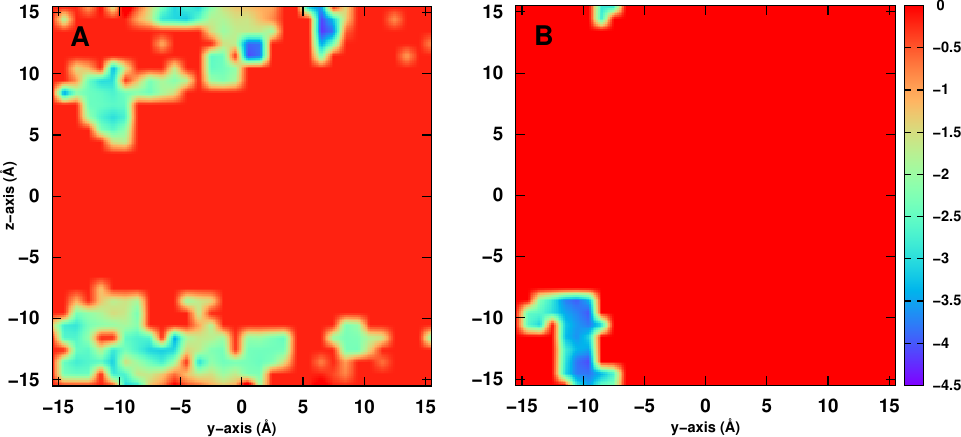}
\includegraphics[scale=0.70]{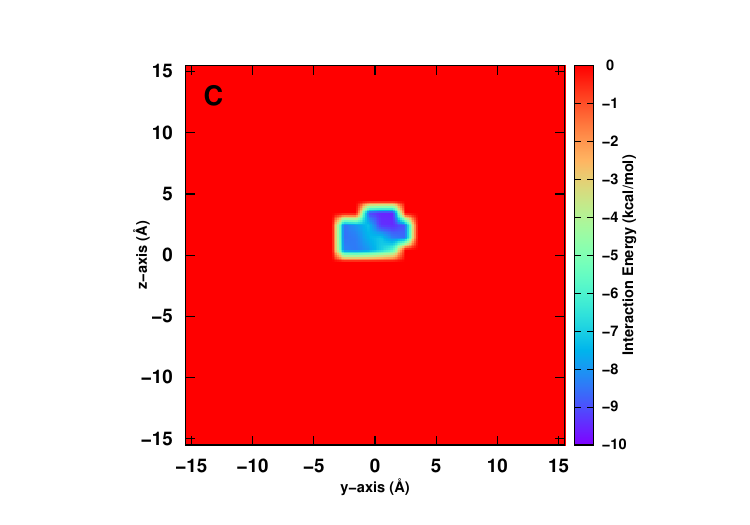}
\caption{Interaction energy between the respective adsorbate and the
  ASW projected onto the $y/z-$plane from a 30 ns long simulation for
  NO (A), CO$_2$ (B), and NO$_2$ (C) on ASW at 50 K. Note that the
  color codes in panels A and B are the same but differ from that in
  panel C.}
\label{sifig:diffusion-50k}
\end{figure}

\begin{figure}[h!]

 \centering \includegraphics[scale=0.7]{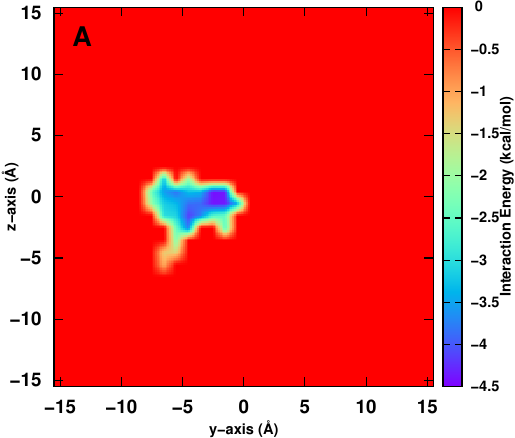}
 \includegraphics[scale=0.7]{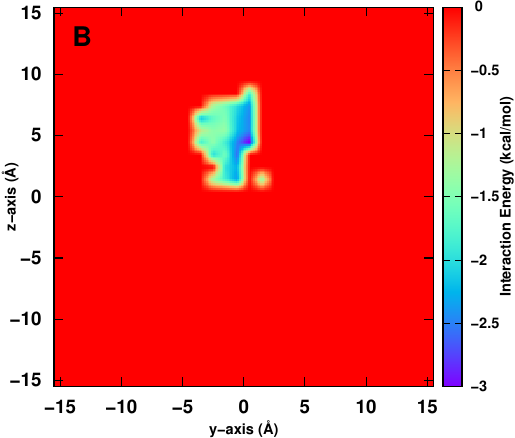}
    \caption{Diffusion of CO$_2$ (Panel A) and NO$_2$ (Panel B) on the
      water surface after recombination (XO+O $\longrightarrow$
      XO$_2$) was observed in simulations running for 10 ns and 6 ns,
      respectively. This allows for comparison with the diffusion at
      50 K, as shown in Figure \ref{sifig:diffusion-50k}.}
    \label{sifig:diffusion-simulations}   
\end{figure}